\documentclass[fleqn,usenatbib]{mnras}

\usepackage{newtxtext,newtxmath}

\usepackage[T1]{fontenc}
\usepackage{ae,aecompl}
\usepackage{supertabular}
\usepackage{appendix}
\usepackage{amsfonts}

\usepackage{graphicx}	
\usepackage{amsmath}	
\usepackage{amssymb}	
\usepackage{longtable}
\usepackage{tabularx}
\defcitealias{2001MNRAS.328..527D}{DM}
\defcitealias{1998AA...330..265L}{LH}

\title[The cooling-down central star of SwSt\,1]{The cooling-down central star of the planetary nebula SwSt\,1: a late thermal pulse in a massive post-AGB star?}

\author[]{Marcin Hajduk,$^{1}$\thanks{E-mail: marcin.hajduk@uwm.edu.pl}
Helge Todt,$^{2}$
Wolf-Rainer Hamann,$^{2}$
\newauthor  Karolina Borek,$^{1}$
Peter A. M. van Hoof,$^{3}$
and Albert A. Zijlstra$^{4}$
\\
$^{1}$Space Radio-Diagnostics Research Centre, University of Warmia and Mazury, ul.Oczapowskiego 2, 10-719 Olsztyn, Poland\\
$^{2}$Institut f\"{u}r Physik und Astronomie, Universit\"{a}t Potsdam, Karl-Liebknecht-Str. 24/25, D-14476 Potsdam, Germany\\
$^{3}$Royal Observatory of Belgium, Ringlaan 3, B-1180 Brussels, Belgium \\
$^{4}$Jodrell Bank Centre for Astrophysics, School of Physics and Astronomy, University of Manchester, Manchester, M13 9PL, UK \\}

\date{Accepted XXX. Received YYY; in original form ZZZ}

\pubyear{2015}

\begin{document}
\label{firstpage}
\pagerange{\pageref{firstpage}--\pageref{lastpage}}
\maketitle

\begin{abstract}
SwSt\,1 (PN\,G001.5-06.7) is a bright and compact planetary nebula containing a late [WC]-type central star. Previous studies suggested that the nebular and stellar lines are slowly changing with time. We studied new and archival optical and ultraviolet spectra of the object. The  [O\,{\sc iii}] 4959 and 5007\,\AA\ to $\mathrm{H}\beta$ line flux ratios decreased between about 1976 and 1997/2015. The stellar spectrum also shows changes between these epochs. We modeled the stellar and nebular spectra observed at different epochs. The analyses indicate a drop of the stellar temperature from about 42\,kK to 40.5\,kK between 1976 and 1993. We do not detect significant changes between 1993 and 2015. The observations show that the star performed a loop in the H-R diagram. This is possible when a shell source is activated during its post-AGB evolution. We infer that a late thermal pulse (LTP) experienced by a massive post-AGB star can explain the evolution of the central star. Such a star does not expand significantly as the result of the LTP and does not became a born-again red giant. However, the released energy can remove the tiny H envelope of the star.
\end{abstract}




\section{Introduction} \label{sec:intro} 

Low- and intermediate mass stars start their evolution when H ignites in the core. After hydrogen is exhausted in the core, H burning proceeds to a shell around the He core. The star ascends the red giant branch (RGB). When the helium core ignites, the star enters the He burning phase. After He burning proceeds to the shell, the products of the nucleosynthesis form a C/O core. The star ascends the asymptotic giant branch (AGB). At first, nucleosynthesis proceeds in a H-burning and a He-burning shell. Since He burning is faster, the He shell approaches the H shell. Then burning becomes unstable and thermal pulses occur. 

A pulse driven convective zone (PDCZ) develops in the region between the hydrogen and helium shell (intershell) as the result of a helium flash. The PDCZ does not reach the H-rich envelope. Thus, hydrogen remains the most abundant element in the atmosphere after the star leaves the AGB in this scenario. However, the surface abundances are enhanced in He and C synthesized in helium shell flashes on the AGB. The dredge up occurs when the convective envelope reaches the layers where nuclear processing takes place \citep{2004agbs.book....1H}. 

He flash events continue while the envelope of the star is being consumed and ejected due to nuclear burning and mass loss \citep{2005ARA&A..43..435H}. After most of the envelope is lost, the core of the star is exposed. The temperature of the star evolves extremely fast. The crossing time $\tau_\mathrm{cross}$ from the end of AGB phase to reach a maximum temperature over $10^5 \, \mathrm{K}$ ranges from several hundreds of years up to several $10^4$ years \citep{2016A&A...588A..25M}. During this period the configuration of the star changes dramatically. The radius shrinks by about 3 orders of magnitude. The harder stellar radiation ionizes previously ejected matter, which form a planetary nebula (PN). When the thermonuclear reactions extinguish, the star descents on the white dwarf (WD) cooling track.

The AGB final thermal pulse (AFTP) can encounter a star with a relatively small envelope mass \citep[see][for review]{2001Ap&SS.275...41K}. As the result, the hydrogen burning ceases. The hydrogen atmosphere of the star is mixed with the intershell as the result of the dredge up. However, the hydrogen abundance at the surface of the star is not significantly red diluted.

A helium flash can also occur during the horizontal evolution of the post-AGB star. Such so-called late thermal pulse (LTP) forces the star to return to the AGB for the second time. The hydrogen mass fraction decreases to a few percent due to dredge up, but is still detectable in the atmosphere. The best known example of an LTP object is FG\,Sge \citep{2006A&A...459..885J}. Another candidate of an LTP is SAO\,244567, the central star of the Stingray Nebula \citep{2017MNRAS.464L..51R}.

The helium layer can re-ignite when the H-burning shell is almost extinguished on the WD cooling track, leading to a very late thermal pulse (VLTP). This phenomenon has been anticipated in theoretical models of stellar evolution published by \citet{1979A&A....79..108S} and \citet{1983ApJ...264..605I}. As the result of this violent event, the star retraces its evolution, becoming a "born-again" AGB star. The remaining hydrogen shell is no longer separated from the helium layer and is consumed by the PDCZ, leaving a hydrogen deficient stellar surface. The released energy causes the first return of the star to the AGB domain. A second return to the AGB domain occurs on a much longer timescale due to a helium flash.

The VLTP phenomenon provides a nice and elegant explanation for the hydrogen-deficient [WC]-type central stars. This link has been strengthed by the discovery of a [WC]-like appearance of the spectrum of V605\,Aql a few decades after its outburst \citep{2006ApJ...646L..69C}.

However, some observational facts contradict [WC] stars being the descendants of the VLTP or the LTP. The proportion of the H-deficient stars among young nebulae is the same as in the whole population \citep{2000A&A...362.1008G}. \citet{2000A&A...362.1008G} argue that [WC] central stars evolve directly from the AGB phase due to the AFTP. However, AFTP would lead to higher H surface abundances than observed in [WC] stars. \citet{2003MNRAS.340..253W,2008MNRAS.383.1639W} found an unexpectedly high Ne abundances and low C/O ratio in the ejecta of "born-again" stars, which are not predicted by VLTP models. Thus, the exact mechanism of the formation of [WC] stars remains unclear.

One of the most robust ways to study the evolution of post-AGB stars is following the real time changes in the stellar and nebular spectra over long time intervals. When high quality stellar spectra are available, atmosphere modeling can reveal the evolution of the stellar parameters in time. A change of the stellar temperature leads to a change of the excitation of the PN in time. Quantitative determination of the heating rate of the central star $\dot{T}_\mathrm{eff}$ provides an important constraint to test and refine evolutionary models in the transition between the AGB and WD, which one of the phases in stellar evolution that are least understood \citep{2019IAUS..343...36M}. 

We studied the changing excitation of PNe in \citet{2015A&A...573A..65H} using the [O\,{\sc iii}] 5007\,\AA\ line fluxes. Unexpectedly, we observed a decrease of the excitation with time in one case - PN SwSt\,1, at odds with other central stars. The rest of PNe showed no change or an increase of the excitation, in line with the expected ''regular'' evolution of their central stars.

SwSt\,1 is a compact (1$.\!\!^{\prime\prime}$3) and young planetary nebula. The nebula is hydrogen rich and indicates the kinematical age of only 290 years for the assumed distance of 2\,kpc \citep[][hereafter \citetalias{2001MNRAS.328..527D}]{2001MNRAS.328..527D}. The PN contains both crystalline silicates and carbon dust \citep{2001Ap&SS.275..113S}, and belongs to a class of dual dust chemistry PNe \citep{2009A&A...495L...5P}.

PN SwSt\,1 contains a [WC\,9/10] central star \citep{2002AJ....124..464D}. The star has weaker lines than for the [WC\,9] type but is hotter than [WC\,10].

\citet{1979VA.....23..213C} noted the possible changes of the spectrum of the central star and the nebula SwSt\,1 with respect to previous observations. They claimed a weakening of P Cygni absorption, suggesting decreasing mass loss rate. Later \citet{1987MNRAS.227..773D} noticed that [Fe\,{\sc ii}] lines weakened in time compared to \citet{1979VA.....23..213C}.

\citet{1979VA.....23..213C} reported a diminution of P\,Cyg profiles and disappearance of several weak C\,{\sc iii} and O\,{\sc iii} lines between 1940 and 1962 reported by \citet{1979VA.....23..213C}. \citetalias{2001MNRAS.328..527D} reexamined old observations and did not support this.

However, \citetalias{2001MNRAS.328..527D} noted that the C\,{\sc iv} $ \lambda 5471 $ line observed by \citet{1977JRASC..71...67A} is much stronger and broader than in their spectrum. C\,{\sc iv} 5801 and 5812\,\AA\ lines did not show much change. They suggest that [Fe\,{\sc ii}] lines may have weakened, which they attribute to possible increase of the excitation of the PN.

\section{Observations} \label{sec:observations}

\begin{table*}
\renewcommand{\thetable}{\arabic{table}}
\centering
\caption{List of observations of SwSt\,1.} \label{tab:observations}
\begin{tabular}{ccccccc}
\hline
\hline
\\[-1.5ex]
Wavelength range [\AA] & Telescope & Instrument & Resolution & Date & Aperture [\arcsec] & Exposure time [s] \\
\\[-1.5ex]
\hline
\\[-1.5ex]
1150-1978 & IUE & SWP  & 200-400& 1979-06-19 & $20 \times 10$ &3600  \\
1851-3348 & IUE & LWR  & 200-400& 1980-09-07 & $20 \times 10$ &1200  \\
1150-1974 & IUE & SWP  & 10000 & 1982-05-30 & $20 \times 10$ &7800  \\
1140-1728 & HST & STIS & 45800 & 2000-09-05 & $0.2 \times 0.2$ &2303 \\
1587-3172 & HST & STIS & 415-730& 2000-09-27 & $0.2 \times 0.2$&2167 \\
3537-3819 & HST & STIS & 415-730& 2000-10-01 & $52 \times 0.1$ &100  \\
5450-6014 & HST & STIS & 425-680& 2000-10-01 & $52 \times 0.1$ &150  \\
5996-6531 & HST & STIS & 425-680& 2000-10-01 & $52 \times 0.1$ &150  \\
6483-7051 & HST & STIS & 425-680& 2000-10-01 & $52 \times 0.1$ &195  \\
6995-7563 & HST & STIS & 425-680& 2000-10-01 & $52 \times 0.1$ &70   \\
904-1188  &FUSE & FUV  & 15000 & 2001-08-21 & $30 \times 30$ &23141 \\
904-1188  &FUSE & FUV  & 15000 & 2001-08-22 & $30 \times 30$ &15331 \\
3722-5572 &SALT & HRS  & 65000 & 2015-05-03 & 1.6	& 1740 \\
5440-8814 &SALT & HRS  & 65000 & 2015-05-03 & 1.6	& 1740 \\
3820-9006 &Mercator&HERMES& 85000 & 2015-05-18 & 2.3 & 1200 \\
3284-4520 & VLT & UVES & 60000 & 2015-07-05 & $8 \times 0.6$ &2420  \\
4619-5599 & VLT & UVES & 70000 & 2015-07-05 & $8 \times 0.6$ &2420  \\
5668-6645 & VLT & UVES & 70000 & 2015-07-05 & $8 \times 0.6$ &2420  \\
\\[-1.5ex]
\hline
\end{tabular}
\end{table*}

\subsection{New optical spectra}

We observed PN SwSt\,1 with UVES mounted on the Very Large Telescope (VLT). Two 1200\,s exposures were supplemented by two 10\,s spectra. The short exposures avoid saturation of the strongest nebular emission lines, while long exposures reveal fainter lines. The DIMM seeing varied between 1{\farcs}01 and 1{\farcs}85 throughout the observations. The configuration of the instrument is presented in Table \ref{tab:observations}. We reduced the spectra with the European Southern Observatory (ESO) Reflex environment \citep{2013A&A...559A..96F}. The spectra were flux calibrated using the spectrophotometric standard Feige\,110, observed in the same night.

We measured the nebular emission lines with a gaussian fit. A systematic difference exists between the absolute fluxes derived from the short and long exposures due to the variable atmospheric conditions. We determined a constant factor of 1.06 by comparing the fluxes derived from the same set of lines measured in both spectra. Then we applied this constant to those fluxes obtained from the short exposure which were saturated in long exposure spectra. In the final list we combined the fluxes of faint lines determined from the long exposure spectrum and the fluxes of strong lines determined from the short exposure spectrum multiplied by the constant factor.

In addition, we took a 1200\,s spectrum with the fibre fed High-Efficiency and High-Resolution Mercator Echelle Spectrograph (HERMES) mounted on the Mercator telescope \citep{2011A&A...526A..69R} on 2015 May 18. The spectrum was reduced by the automatic reduction pipeline. We normalized the spectrum to the assumed stellar continuum and applied radial velocity correction.

The Southern African Large Telescope (SALT) observed SwSt\,1 on May 3, 2015, using fibre feed High Resolution Spectrograph. We used slow readout speed, for both detectors, without binning. Two 870\,s exposures and two 10\,s exposures were performed simultaneously in the blue arm and red arm in high resolution mode. The data processing comprised of bias correction, flat fielding, wavelength calibration, and sky subtraction. We normalized the spectra to continuum and applied radial velocity correction.

\subsection{Archival spectra}

We used spectra from the International Ultraviolet Explorer (IUE) with large aperture, since small aperture spectra show light loss. The spectra with large aperture were collected on 1979 June 18, 1980 September 06, and 1981 March 06. High resolution spectrum was collected on 1982 May 30.

We also retrieved archival Hubble Space Telecope (HST) and Far Ultraviolet Spectroscopic Explorer (FUSE) spectra. All the observations are summarized in Table\,{\ref{tab:observations}}.

\citet{1977JRASC..71...67A} took the spectrum of the central star of PN SwSt\,1 with the Lick 3-metre telescope with an image-tube scanner before 1977. The exact date of the spectra is not given in the paper.
We have digitized the spectra from the scanned publication. All the observations have been normalized by eye to the assumed stellar contiuum.

\subsection{Imaging}

\begin{figure}
\includegraphics[width=1.0\columnwidth]{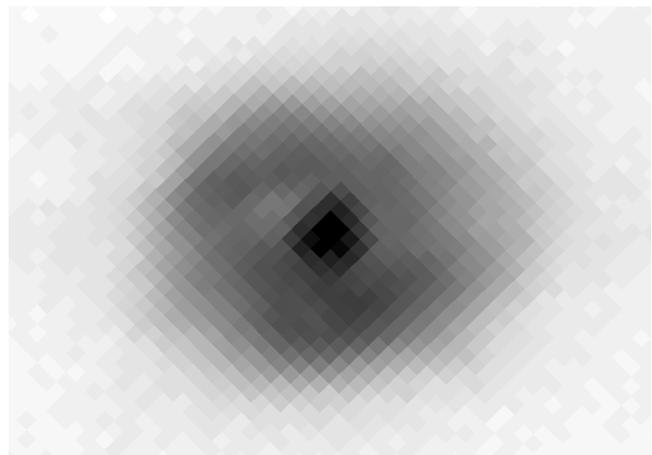} \caption{The HST/MIRVIS image of the PN SwSt\,1. North is on the top, east is on the left. Negative log intensity scale is applied. The image size is $2{\farcs}35 \times 1{\farcs}7$.}
\label{fig:mirvis}
\end{figure}

In addition to optical spectra, the HST obtained images using MIRVIS with 2.1\,s exposure and long pass filter on October 1, 2000. The images were taken along with the STIS spectra. The image shows a projected ellipsoidal shell at a position angle of $90^\circ$. The outer radii are $\rm 0{\farcs}65 \times 0{\farcs}75$ at 10\% of intensity and inner radius of about 0{\farcs}2 (Figure \ref{fig:mirvis}). The nebula also appear to show bipolar outflow at the same PA. The southern part of the nebula is brighter.

\section{Stellar spectral analysis}

\begin{table}
\renewcommand{\thetable}{\arabic{table}}
\centering
\caption{Basic parameters adopted and derived from atmosphere modelling of the central star of the PN SwSt\,1.} \label{tab:powr}
\begin{tabular}{ccccc}
\hline
\hline 
\\[-1.5ex]
 & \citetalias{1998AA...330..265L}$^a$ & \citetalias{2001MNRAS.328..527D}$^b$ & \citetalias{2001MNRAS.328..527D}$^c$ & This paper \\
 \\ [-1.5ex]
 \hline
\\[-1.5ex]
$T_* \rm [kK]$ & $35$ & $40 \pm 2$ & $46$ &  $41.9 \pm 0.5$ \\ 
$\log(L_* / \mathrm{L}_{\odot})$ & 3.27 & 3.95 & 3.95 & 3.27 \\
$\log(\dot{M}/(\mathrm{M}_{\odot}\,\mathrm{yr}^{-1}))$ & $-6.90$ & $-7.02^d$ & $-7.16^d$ & $-6.69^d$ \\ 
$\varv_{\infty}$ [km/s]& 400 & 900 & 800 & 800 \\
$d$\,[kpc] & 1.4 & 2 & 2 & 2 \\
$E(B-V)$ & 0.41 & 0.46 & 0.48 & 0.32 \\
$\rm \beta_\mathrm{H}$ & $\leq 10$ & unknown & 0 & $< 5$ \\  
$\rm \beta_\mathrm{He}$ & 44 & 37 & 53 & 42 \\
$\rm \beta_\mathrm{C}$ & 53 & 51 & 32 & 50 \\
$\rm \beta_\mathrm{O}$ & 3 & 12 & 15 & 3 \\
$\rm \beta_\mathrm{N}$ & $<0.5$ & ${\sim}0$ & 0 & 0.05 \\
$\rm \beta_\mathrm{Si}$ & $\approx 1$ & & & 0.1 \\ 
$\rm \beta_\mathrm{Ne}$ &2..4 & & & 0.2 \\ 
\\[-1.5ex]
\hline
\end{tabular}

$^a$ \citet{1998AA...330..265L}
$^b$ {\sc isa} -- WIND model \citep{2001MNRAS.328..527D}
$^c$ {\sc cmfgen} model \citep{2001MNRAS.328..527D}
$^d$ obtained with the clumping parameter of 10
\end{table}

\begin{figure}
\includegraphics[angle=270,width=\columnwidth]{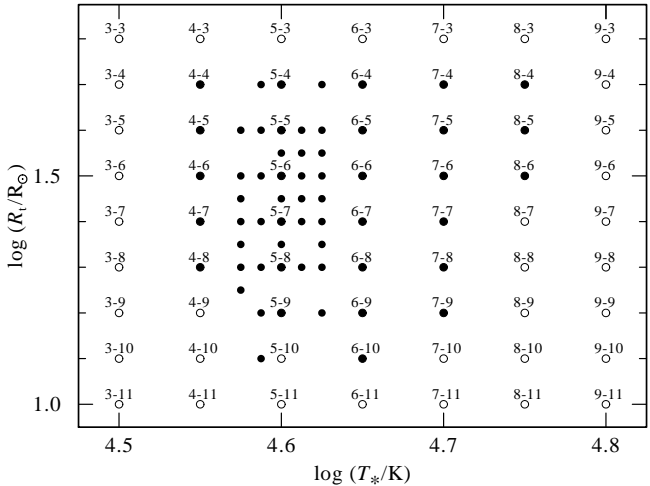} \caption{PoWR models computed in the $\log R_\mathrm{t} - \log T$ grid (filled circles). Each model is marked with $k-m$ notation (Eqs.\,\ref{eq:3} and \ref{eq:4}).}
\label{fig:powr}
\end{figure}

\begin{figure*}
\includegraphics[angle=270,width=\textwidth]{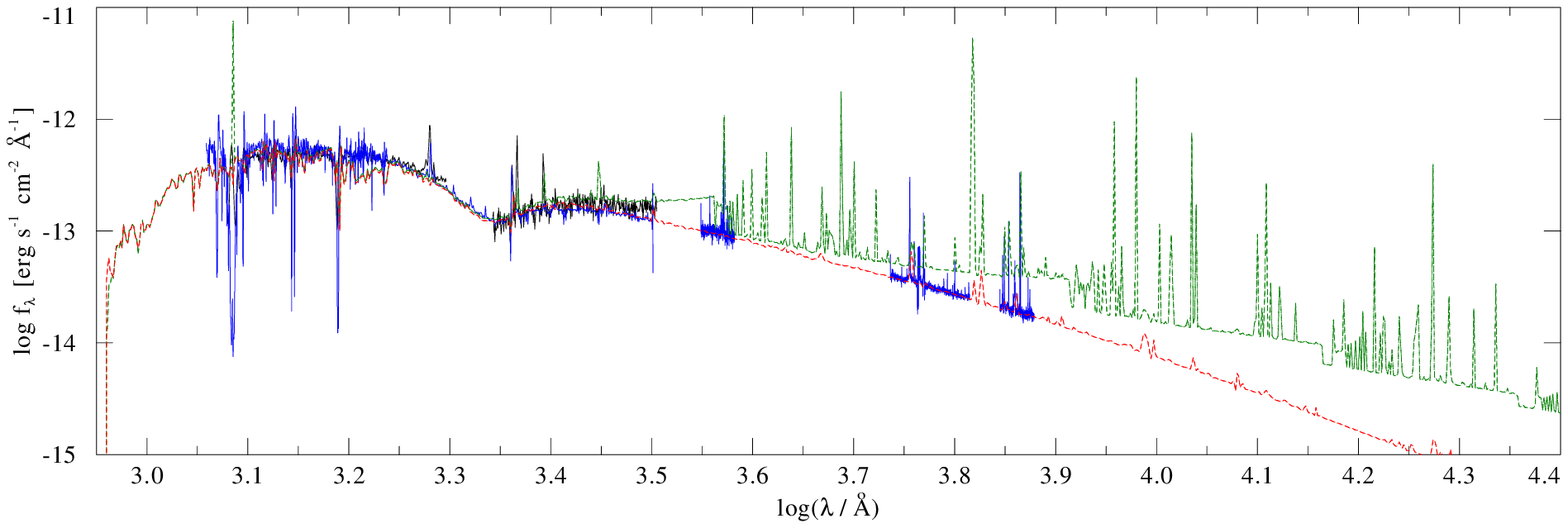} \caption{Observed and modeled SED of SwSt\,1. The red line shows the PoWR model spectrum of the central star (Table\,\ref{tab:powr}). The dark-green line contains the Cloudy spectrum of the PN SwSt\,1 (Table\,\ref{tab:cloudy}). The synthetic spectra are scaled to the distance of 2\,kpc and reddened with  $E(B-V)=0.32 \, \mathrm{mag}$ using the extinction law by \citet{1999PASP..111...63F} with the parameter $R_V=3.1$. Black lines represent IUE observations. Blue lines represent HST observations. Note that aperture for the HST spectra was much smaller than for the IUE spectra, so the contribution of the nebular lines and continuum is much smaller (Tab.\,\ref{tab:observations}). The FUSE spectra were not included in the fit. They show an excess in the blue with respect to the other spectra, most likely due to calibration uncertainties.}
\label{fig:cloudy}
\end{figure*}

We calculate atmosphere models of the central star of SwSt\,1 with the Potsdam Wolf-Rayet (PoWR) code described in \citet{2002A&A...387..244G} and \citet{2004A&A...427..697H}. The code calculates non-LTE radiative transfer for a spherically symmetric and stationary outflow with a predescribed velocity field. The stellar radius $R_*$ defines the inner boundary of the model, with the radial optical depth of $\tau_\mathrm{Ross} = 20$.

Stellar temperature $T_*$, radius $R_*$, and luminosity $L_*$ are related by Stefan-Boltzmann's law:
\begin{equation}
L_* = 4 \pi R_*^2 \sigma_{\mathrm{SB}} T_* ^4
\end{equation}

\citet{1989A&A...210..236S} showed that Wolf-Rayet type model spectra are invariant with respect to different combinations of mass loss $\dot{M}$ and $R_*$ parameters. In first approximation, the strength of the emission lines depends only on the \emph{transformed radius}

\begin{equation}
R_\mathrm{t} = R_* \left[ \frac{\varv_{\infty}}{2500\, \mathrm{km}\,\mathrm{s^{-1}}} \middle/ \frac{\dot{M} \sqrt{D}}{10^{-4}\,\mathrm{M}_{\odot} \mathrm{yr}^{-1}} \right]^{2/3}.
\label{eq:rt}
\end{equation}

where $\varv_{\infty}$ denotes the terminal velocity of the stellar wind and $D$ is the clump density contrast (the ratio of the density of clumps to the average density).

\subsection{Temperature and transformed radius}

We computed a grid of models in a $\log(R_\mathrm{t}) - \log(T_{\star})$ parameter space (Fig.\,\ref{fig:powr}). We set the luminosity to $\log(L_* / \mathrm{L}_{\odot}) = 3.27$. We did not attempt to match the exact value of $L_*$. The models can be later scaled to a different luminosity by a scaling of the mass-loss rate as $\dot{M} \propto L_*^{3/4}$ \citep{2012A&A...540A.144S}.

We adopted, for convenience, the model indexing introduced by \citet{2004A&A...427..697H}:

\begin{equation}
\log (T_*/\mathrm{K}) = 4.35 + 0.05 \, k
\label{eq:3}
\end{equation}

\begin{equation}
\log (R_\mathrm{t}/\mathrm{R}_{\odot}) = 2.1 - 0.1 \, m
\label{eq:4}
\end{equation}
where $k$ and $m$ are integer numbers. We attempted to fit a relative change in the stellar temperature between different dates. These changes are relatively small, comparable to one step in the index $k$. We used a finer grid spacing with $\Delta k$ of 0.5 and 0.25 and $\Delta m$ of 0.5 when necessary.

We fitted the normalized line spectrum in the spectral region of 900--9000\,\AA. Then, when we achieved a reasonable fit, we also fitted the spectral energy distribution (SED). We used spectra calibrated in absolute fluxes for this purpose. 

The stellar SED is well fitted with a color excess of $E(B-V)=0.32 \, \mathrm{mag}$ (Fig.\,\ref{fig:cloudy}). The reddened spectrum fits well the observed dip in the SED at about 2200\,\AA\ caused by interstellar extinction. The nebular $ \mathrm{H}\alpha / \mathrm{H}\beta $ ratio dereddened with $E(B-V) = 0.32 \, \mathrm{mag}$ is close to the theoretical ratio.

The IUE long-wavelength spectrum does not agree with the HST spectrum. This may be caused by the contribution of the nebular bound-free emission (Fig.\,\ref{fig:cloudy}), as the aperture of IUE was significantly larger than that of HST. HST spectra do not contain a lot of contribution from the nebula.

\begin{figure*}
\includegraphics[width=0.7\textwidth, angle=0]{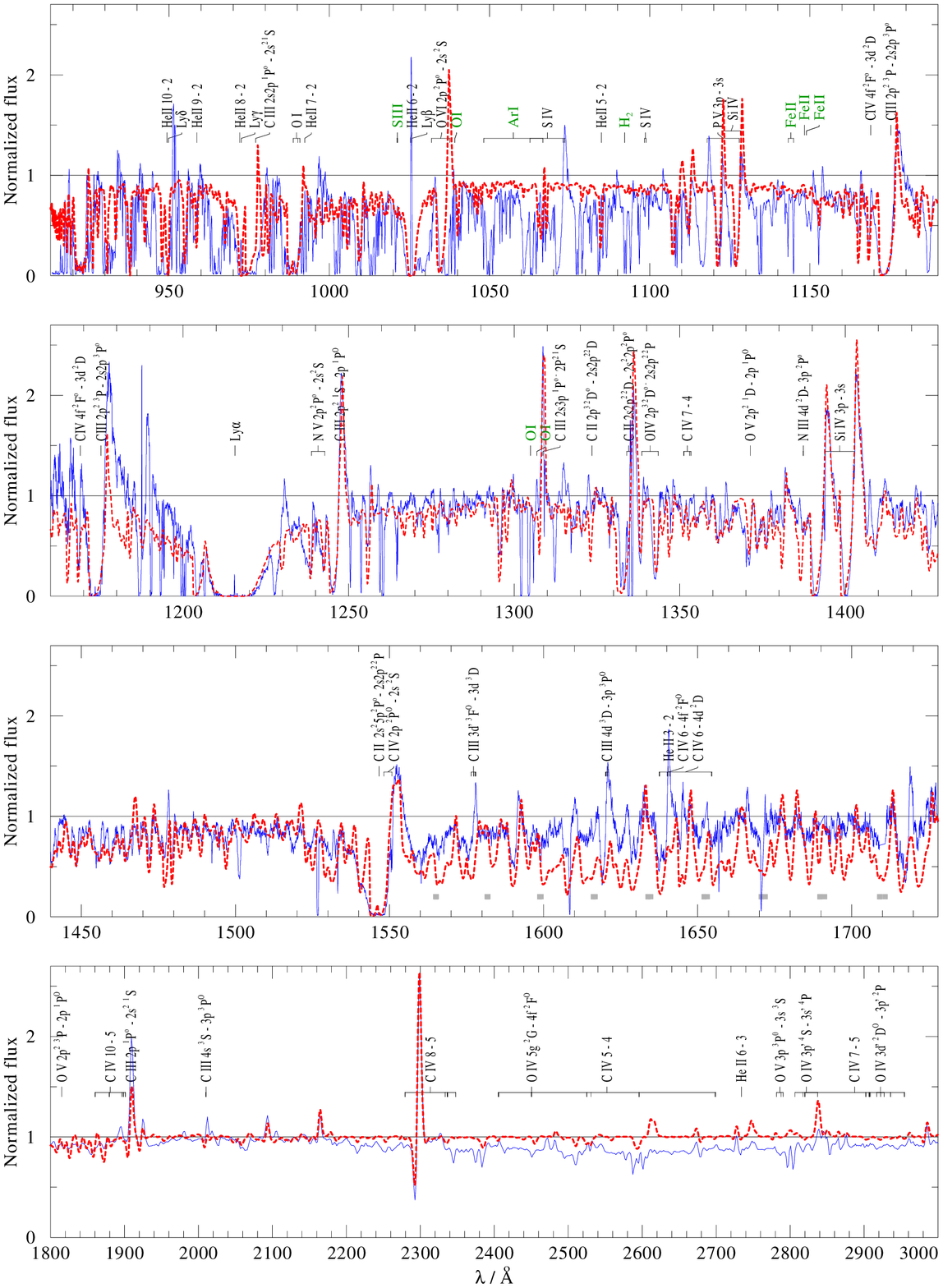}
\caption{The UV spectrum of SwSt1: Observations (FUSE and HST STIS, see Table\,\ref{tab:observations}) are shown with a thin blue line. The best fitting PoWR model is plotted with a thick red dashed line. The HST spectra are normalized by the model continuum, while the FUSE spectrum is normalized by eye.  In the wavelength range of 913-1800\,\AA{} observations and model are smoothed with a Gaussian of 0.1\,\AA{} FWHM for a better representation. For the range of 1800-3000\,\AA{} the model is convolved with a Gaussian of 5\,\AA{} FWHM to match the resolution of the observation. Gaps in the observed spectrum are indicated by thick gray bars. Non-stellar features are indicated by green lables.}
\label{fig:uv}
\end{figure*}

\begin{figure*}
\includegraphics[width=0.8\textwidth]{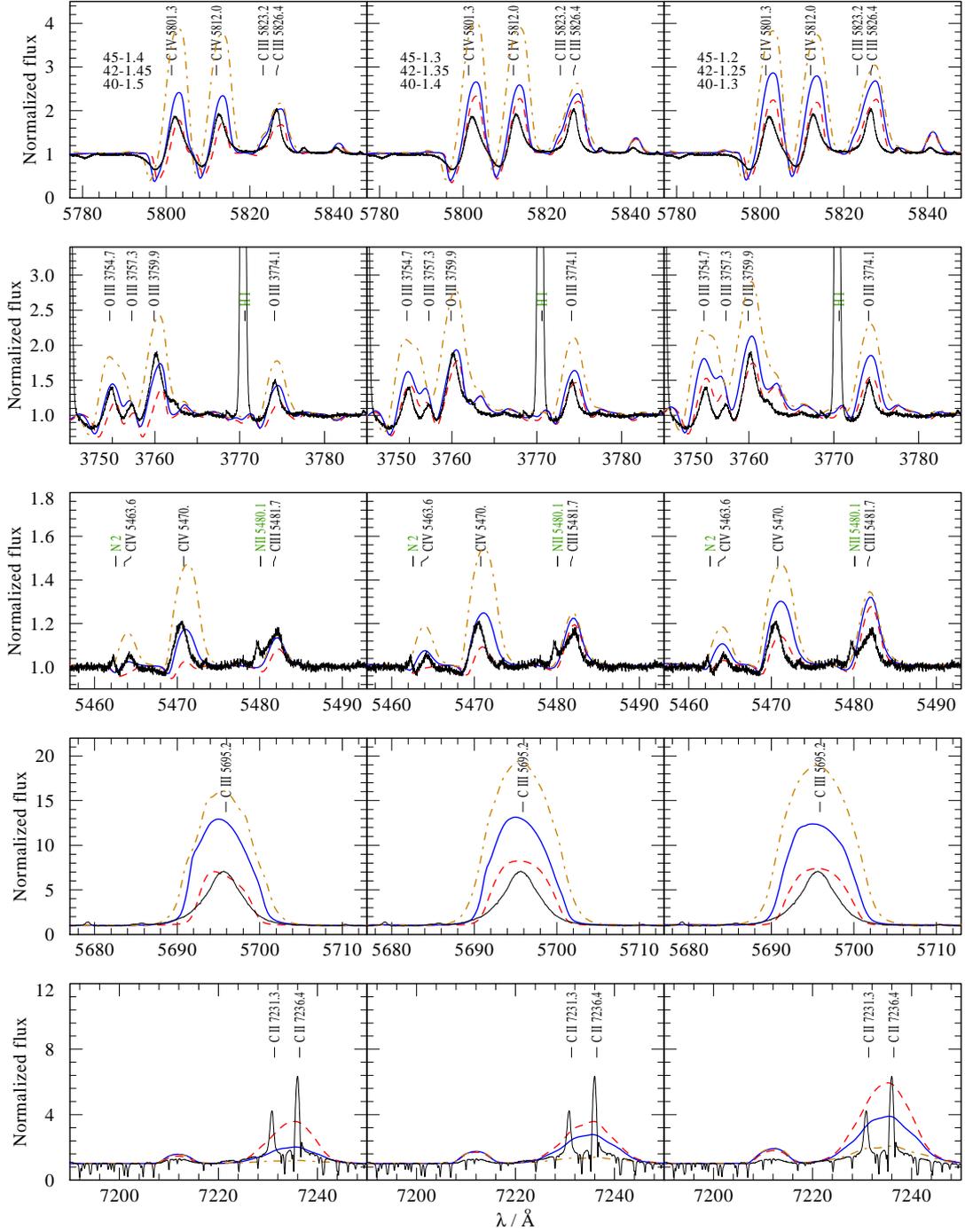}
\caption{PoWR fit to the temperature sensitive lines. The black line represents the VLT spectra (four top rows) and the SALT spectrum (the last row). Each column contains three similar models fitted to the spectrum. The corresponding temperatures in kK and logs of transformed radii in $\mathrm{R}_{\odot}$ for each column are shown in the upper row. Brown dash-dotted lines represent models with $T_* = 45 \, \mathrm{kK}$, blue lines represent models with $T_* = 42 \, \mathrm{kK}$, and red dashed lines models with $T_* = 40 \, \mathrm{kK}$. Different rows show different wavelength ranges. The $R_\mathrm{t}$ is decreasing by 0.1 in log from left to right column. Identifications of nebular lines are in green and stellar in black lines.}
\label{fig:temp}
\end{figure*}

\begin{figure*}
\includegraphics[angle=0,width=0.8\textwidth]{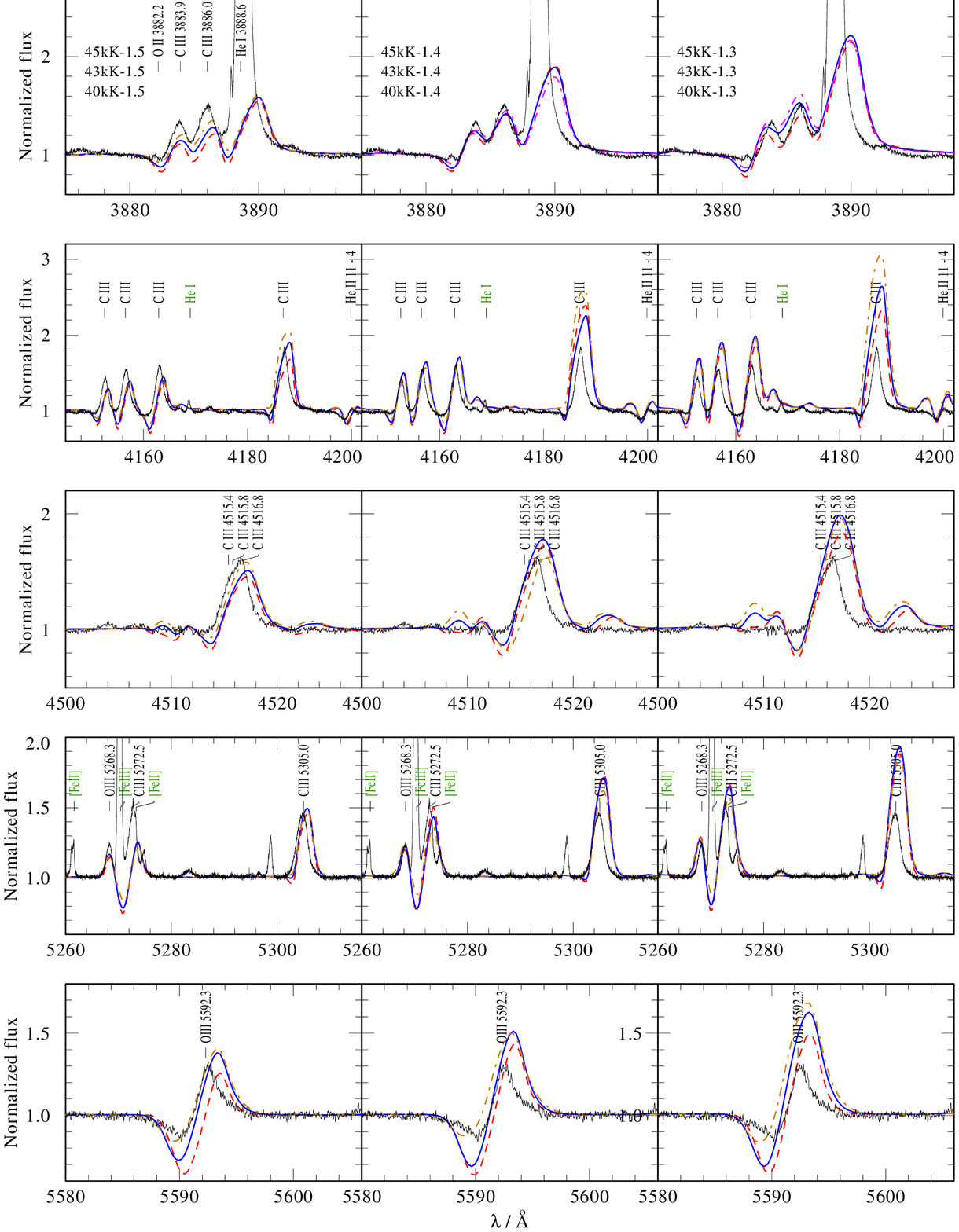}
\caption{PoWR fit to the mass loss sensitive lines. The black line represents the VLT spectra. Each column presents three similar models fitted to the spectrum. The corresponding temperatures in kK and logs of transformed radii in $\mathrm{R}_{\odot}$ for each column are shown in the upper row. Brown dash-dotted lines represent models with $T_* = 45 \, \mathrm{kK}$, blue lines represent models with $T_* = 42 \, \mathrm{kK}$, and red dashed lines -- models with $T_* = 40 \, \mathrm{kK}$. Different rows show different wavelength range. The $R_\mathrm{t}$ is decreasing by 0.1 in log from left to right column.}
\label{fig:rt}
\end{figure*}

Ultraviolet (resonant) lines give information on the terminal velocity (Fig.\,\ref{fig:uv}). \citet[][hereafter \citetalias{1998AA...330..265L}]{1998AA...330..265L} used only optical lines in their analysis, and that is why they underestimated the terminal velocity by a factor of about 2. In optical spectra, the absorption component in the P-Cygni profiles extends only to a velocity of $-400\, \mathrm{km/s}$, while in the UV these features indicate a maximum blueshift of $-800\,\mathrm{km/s}$, indicating that
the optical emission lines form closer to the stellar surface.

For the velocity field $v(r)$ we adopted the so-called two-$\beta$-law \citep{1999ApJ...519..354H}, which is a sum of two beta-laws with exponents $\beta_1$ and $\beta_2$, respectively. We used the exponents $\beta_1 = 1$, $\beta_2 = 8$, and $\varv_\infty = 800\, \mathrm{km} \, \mathrm{s}^{-1}$ distributed between these two terms in the ratio 60:40. These parameters of the velocity field reproduced well the line profiles (see Figure \ref{fig:uv}. We assumed a clumping parameter of 10, i.e. the density in the clumps is 10 times higher than the average density, while the interclump space is void. Such clumping properties have been established for other [WC] stars as well (Todt et al. 2008).

We searched for the best fit of our spectra in the $\log(R_\mathrm{t}) - \log(T_{\star})$ parameter space. At the beginning we searched for a set of emission lines, which are sensitive to stellar temperature and only weakly depend on transformed radius. Each column in Fig.\,\ref{fig:temp} shows three models with different temperatures and transformed radii to different spectral regions. $R_\mathrm{t}$ is decreasing by 0.1 from left to right column, while $T_*$ remain the same. It can be noted, that models with different temperatures give different line strengths. However, the line strengths are roughly invariant with respect to $R_\mathrm{t}$, since the line strengths do not change much between different columns. Thus, the stellar lines shown in Fig.\,\ref{fig:temp} depend mainly on the stellar temperature.

Models with $T_* = 45\,\mathrm{kK}$ overestimate the strength of the most of the stellar lines. The best fit is achieved for $T_* = 40-41\,\mathrm{kK}$. The temperature sensitive line C\,{\sc iv} 5475\,\AA\ indicates model with $T_* = 41 \, \mathrm{kK}$, while C\,{\sc iv} 5801/5812\,\AA\ and C\,{\sc iii} 5696\,\AA\ favour model with $T_* = 40 \, \mathrm{kK}$.

Then we identified stellar lines which are more sensitive to the value of the transformed radius. Each column in Fig.\,\ref{fig:rt} shows three models with different temperatures and different transformed radii for different spectral regions. $R_\mathrm{t}$ is decreasing by 0.1 in log from left to right column, for the same values of $T_*$.

The three models with the same temperatures give similar line strengths, thus the presented set of lines is not sensitive to the stellar temperature in the range under consideration. The line strengths increase with the increasing $T_*$, so they depend mainly on the transformed radius.

The models give the best fit for $\log R_\mathrm{t} = 1.4$ (middle column) for most of the lines. Thus, model $T_* = 40-41 \,\mathrm{kK}$ and $\log R_\mathrm{t} = 1.4$ gives the best fit.

\subsection{Atmospheric abundances}

\begin{figure}
\includegraphics[width=\columnwidth]{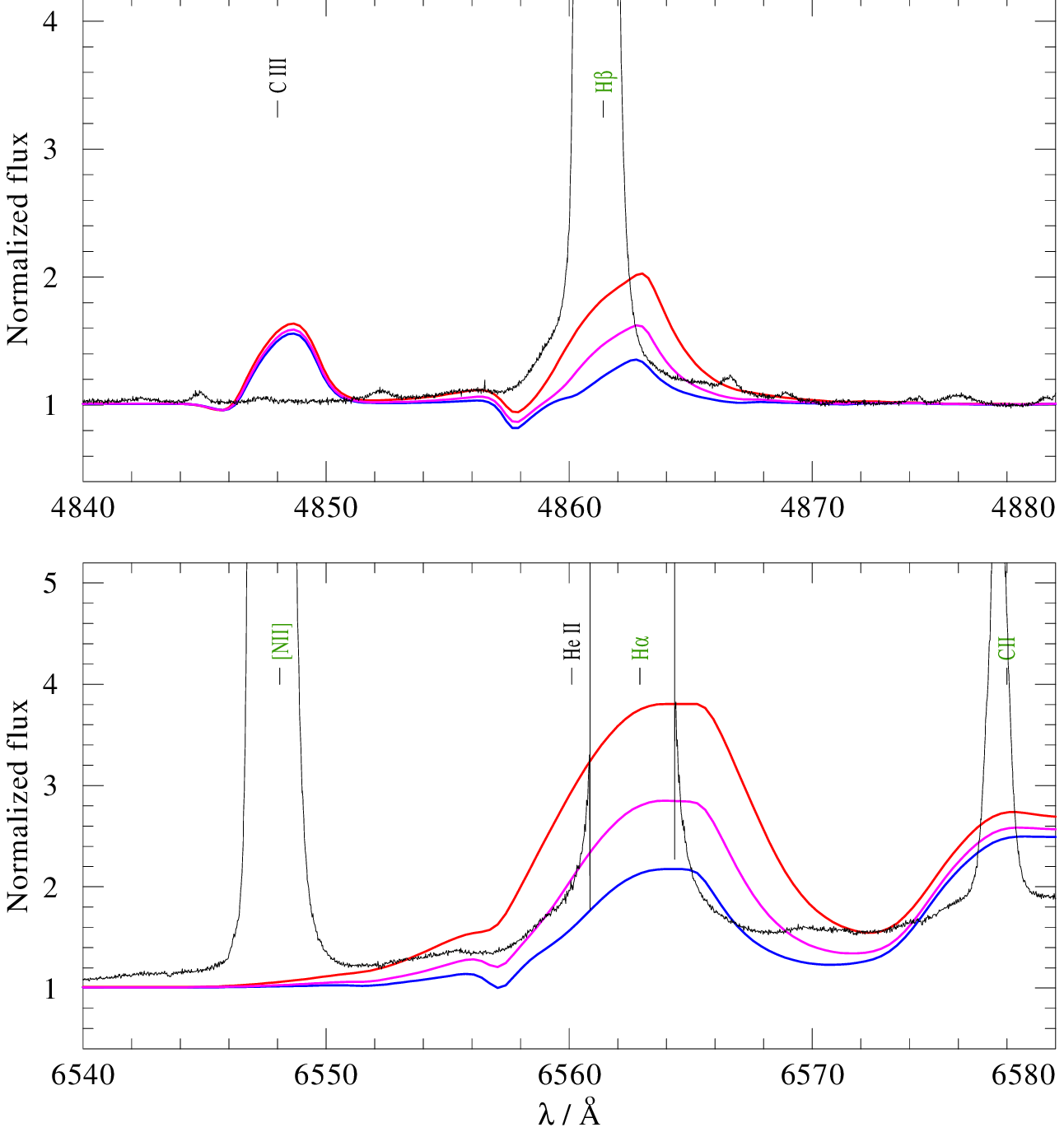}
\caption{PoWR fit in the region of $\rm H\alpha$ and $\rm H\beta$ lines. Different lines correspond to hydrogen abundances of 0.05, 0.1, and 0.2 by mass, respectively. Any larger value than 0.05 can be excluded.}
\label{fig:h}
\end{figure}

In order to optimize the line fit, we also varied the abundances of hydrogen, helium, carbon, oxygen, nitrogen, neon, and silicon (Table\,\ref{tab:powr}), and the iron-group elements. The iron-group elements are treated in the superlevel approximation \citep{2002A&A...387..244G}.

Our modeling indicates approximately 3\% of oxygen by mass. This is much a lower oxygen abundance than given by \citetalias{2001MNRAS.328..527D}, who obtained a mass fraction of 12\% based on the O\,{\sc iii} 5592\,\AA\ line only. Its strength is not very sensitive to the O abundance. The abundance derived by \citetalias{2001MNRAS.328..527D} severely overestimates O\,{\sc ii} line strengths.

We determined 0.05\% of nitrogen by mass using the N\,{\sc iii} 4630.5 and 4634.1\,\AA\ lines, and also weaker lines in other parts of the spectrum.

Fitting to the 4088.8\,\AA\ and 4116.1\,\AA\ Si\,{\sc iv} lines indicated 0.1\% of silicon. This is much lower value than given by \citetalias{1998AA...330..265L}. 

We obtained 0.2\% of neon by mass in the stellar atmosphere using Ne\,{\sc iii} stellar lines present in the spectrum, e.g. at 4097, 4634, 4641\,\AA. \citetalias{1998AA...330..265L} used the neon line at 6403\,\AA\ to derive a much higher atmospheric abundance of this element ($2-4\%$), but we attribute it to the nebular contamination due to better resolution of our spectra.

An upper limit for H is about 5\% by mass (Fig.\,\ref{fig:h}). Higher hydrogen abundances would lead to line wings that would be detected despite of the nebular emission lines.

It can be noted that a C\,{\sc iii} and C\,{\sc ii} lines in the region of $\mathrm{H}\beta$ and $\mathrm{H}\alpha$ are not very well fitted by the model (Fig.\,\ref{fig:h}). However, applying other stellar parameters would result in worse fit in other C\,{\sc iii} and C\,{\sc ii} lines (see e.g. Figs. \ref{fig:temp} and \ref{fig:rt}).

Table \ref{tab:powr} summarizes the resulting stellar parameters and abundances in comparison with \citetalias{2001MNRAS.328..527D} and \citetalias{1998AA...330..265L}.

\subsection{Distance and luminosity}

The dilution of the model flux depends on the distance and the reddening. The synthetic SED was reddened adopting the reddening law from \citet{1999PASP..111...63F}, assuming $R_V = 3.1$ (Fig.\,\ref{fig:cloudy}).

The distances determined using various methods range from 1 to 5 kpc \citepalias{2001MNRAS.328..527D}, with preferred values of 3--5\,kpc. However, \citetalias{2001MNRAS.328..527D} adopted a lower distance of 2\,kpc, since their modeling yielded a luminosity of the central star of $8900\,\mathrm{L}_{\odot}$, close to the high end for a PN central star. Unfortunately, GAIA (second release) gives a negative parallax of $(-3.95 \pm 1.00) \, \rm mas$ \citep{2018A&A...616A...1G}, most likely due to the influence of the nebula. The determined parallax error is much higher than ${\sim}0.04 \, \rm mas$ expected for such a bright star.

We fitted the SED of the observations which are dominated by the stellar component, i.e. FUSE, IUE, and HST. The derived luminosity of $1860\,\mathrm{L}_{\odot}$ for 2\,kpc is substantially below the high end for central stars of PNe. This means that a distance larger than $\mathrm{2\,kpc}$ would be still acceptable for the PN SwSt\,1. Our models are not very luminosity sensitive, so we do not constrain the luminosity (and thus the distance) of the star with our modeling unambiguously. The luminosity scales with $d^2$, and the mass loss rate $\dot{M}$ with $d^{3/2}$ for the same $R_\mathrm{t}$ (cf. Eq.\,\ref{eq:rt}). The parameters of our best fitting model are compiled in Table\,\ref{tab:powr}.

The SED of SwSt\,1 shows prominent dust emission, which is reprocessed stellar radiation. \citetalias{2001MNRAS.328..527D} derived the luminosity re-radiated by dust as $1450\,\mathrm{L}_{\odot}$ for $\rm 2\,kpc$, much lower than the stellar component with $8900\,\mathrm{L}_{\odot}$. This is inconsistent with the results of \citet{1991A&A...250..179Z}, who estimated that only about 50\% of the flux is emitted by the stellar component, and the remaining flux mostly by the dust component. \citet{2015MNRAS.447.1673V} derived $(437 \pm 106) \, \mathrm{L}_{\odot}\, \rm kpc^{-2}$ by modeling the stellar SED, implying a dust luminosity of $1750\,\mathrm{L}_{\odot}$ for $\rm 2\,kpc$, consistent with the luminosity of the dust component derived by \citetalias{2001MNRAS.328..527D}. However, they fitted three components in the infrared and optical.

\subsection{Evolution of the stellar spectrum}

\begin{figure*}
\includegraphics[width=0.65\textwidth]{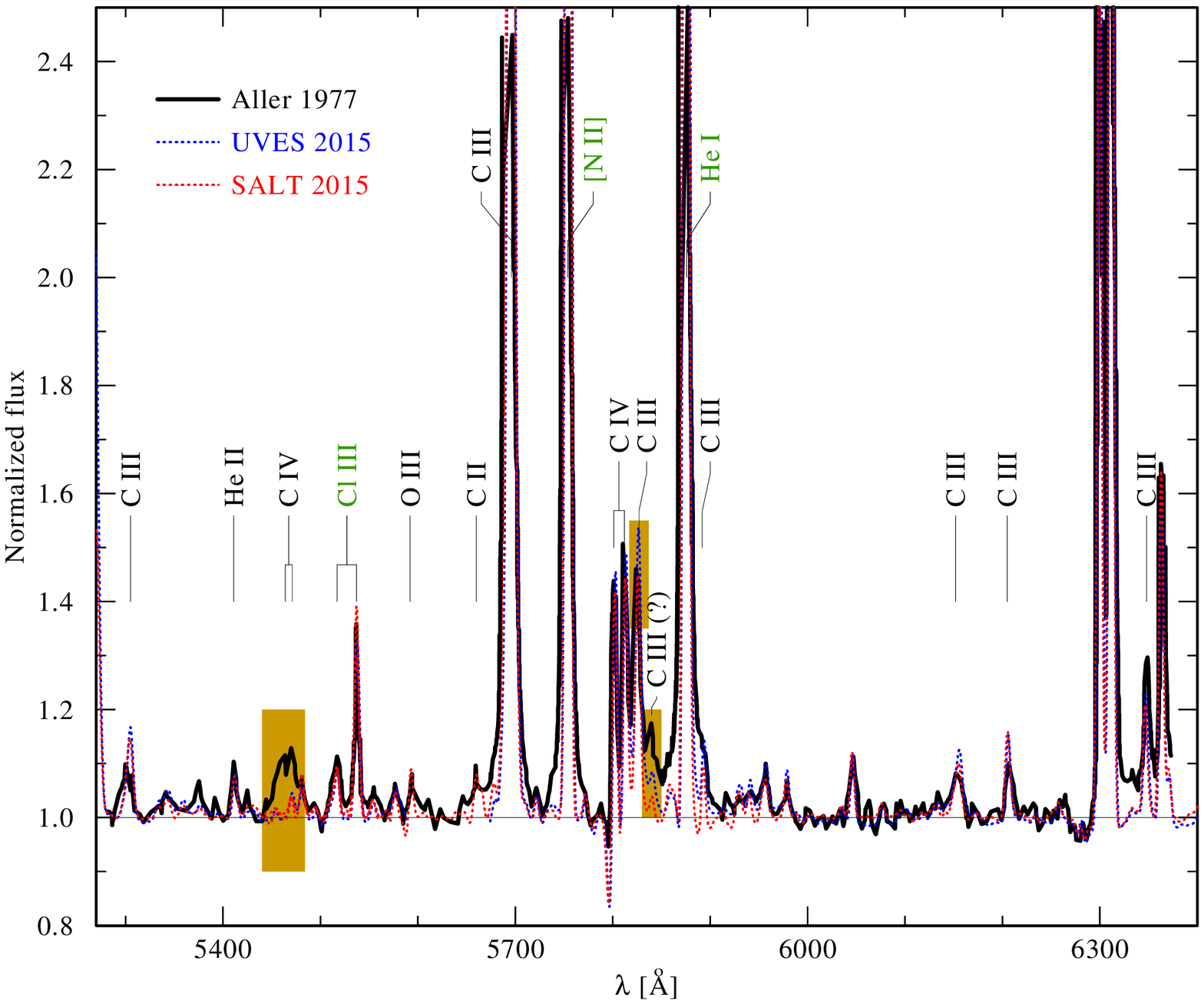} 
\caption{Comparison of observations between different epochs in the
blue-optical range. Black full line: taken before 1977 
at the Lick 3-metre telescope \citep{1977JRASC..71...67A}; blue dotted line: taken 2015 with UVES at the ESO VLT; red dotted line: taken 2015 with the HRS spectrograph at the Southern African Large Telescope (SALT). Nebular lines are identified in green color. The three yellow boxes mark features that seemingly varied with time.}
\label{fig:aller2}
\end{figure*}

\begin{figure*}
\includegraphics[width=0.65\textwidth]{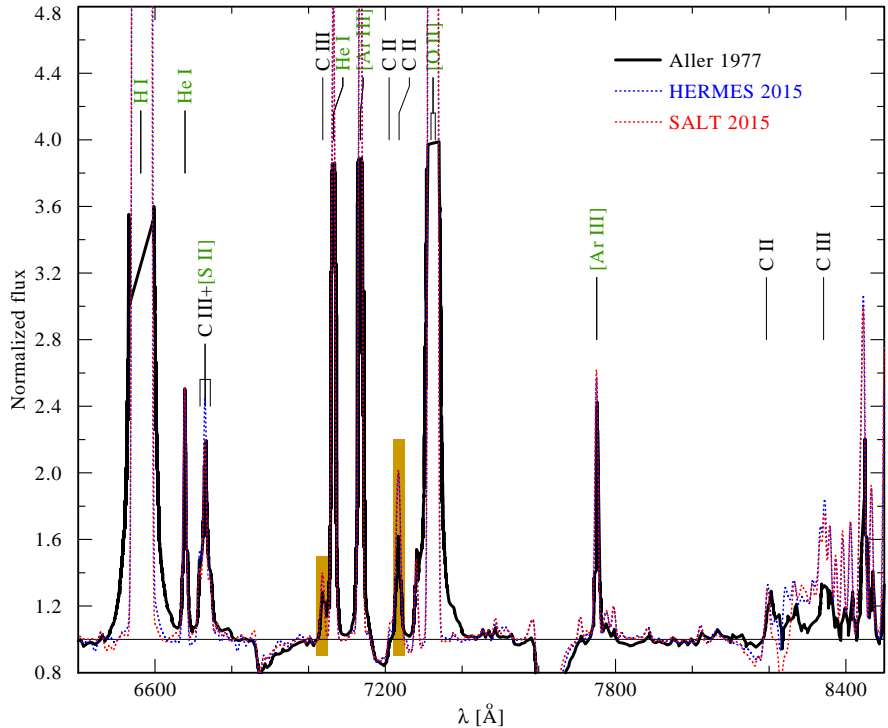} \caption{Comparison of observations between different epochs in the
red-optical range. Black full line: taken before 1977 
at the Lick 3-metre telescope with an image-tube scanner
\citep{1977JRASC..71...67A}; blue dotted line: taken in 2015 with HERMES at the Mercator Telescope; red dotted line: taken 2015 with the HRS spectrograph at the Southern African Large Telescope (SALT). Nebular lines are identified in green color. The two yellow boxes mark
features that seemingly varied with time.}
\label{fig:aller4}
\end{figure*}

\begin{figure*}
\includegraphics[width=0.65\textwidth]{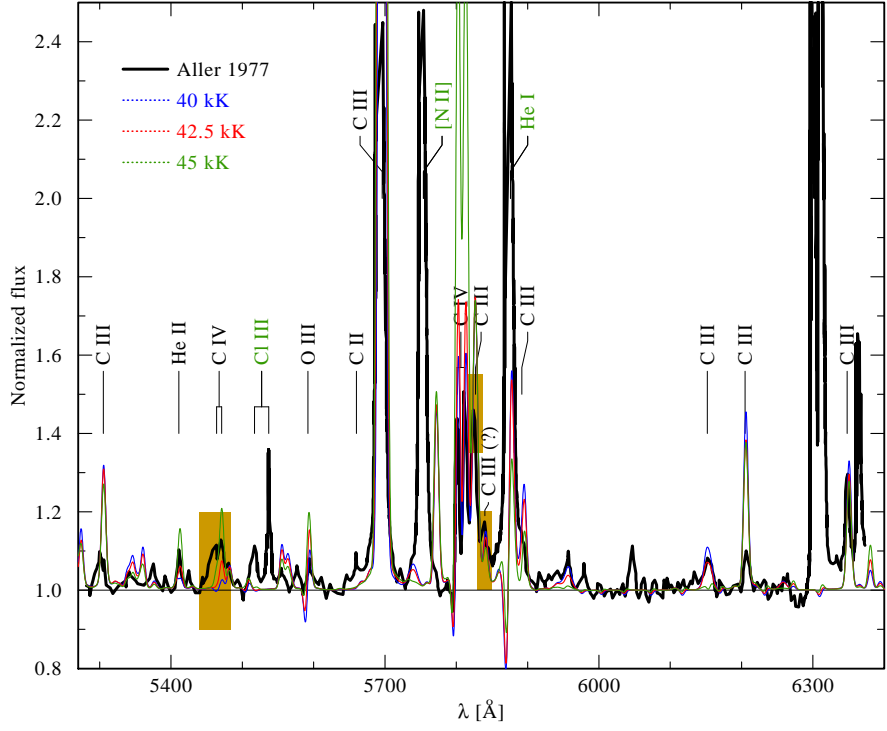}
\caption{Comparison of the blue spectrum from \citet{1977JRASC..71...67A} (black solid line) with the sequence of three PoWR atmosphere models with different temperatures and $\log R_{\mathrm{t}}/\mathrm{R_{\odot}} = 1.4$. Other parameters are given in Table\,\ref{tab:powr}.}
\label{fig:aller1}
\end{figure*}

\begin{figure*}
\includegraphics[width=0.65\textwidth]{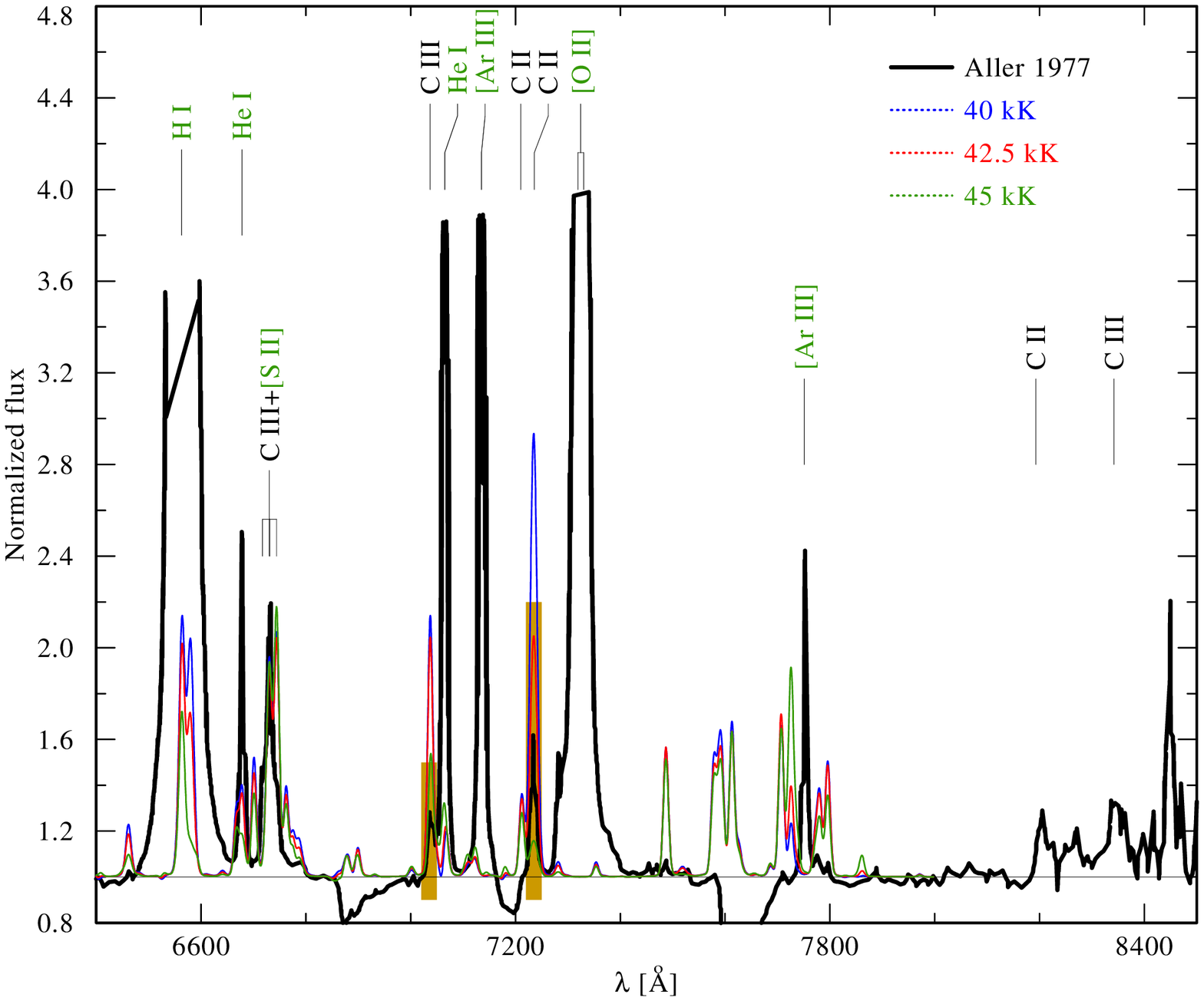} \caption{Same as in Fig.\,\ref{fig:aller1}, but for the red-optical range.}
\label{fig:aller3}
\end{figure*}

Figs.\,\ref{fig:aller2} and \ref{fig:aller4} compare observed spectra of the central star of the planetary nebula SwSt\,1 between 1977 (or earlier) and 2015. The yellow rectangles mark features that seemingly have changed since Aller's observation. Most pronounced is the strong feature at about 5440--5484\,\AA, which became much weaker in the recent observations. The feature is at least partially due to C\,{\sc iv} 5470\,\AA, but combines also unresolved C\,{\sc iii} 5482\,\AA, nebular N\,{\sc ii} 5454\,\AA, and other fainter lines. The nebular lines are very sensitive to the stellar temperature, but their contribution seems to be negligible.

In contrast, the C\,{\sc iii} line at 5827\,\AA\ became stronger with time, now surpassing the two neighboring C\,{\sc iv} lines at 5801 and 5812\,\AA. The trend of weaker C\,{\sc iv} and stronger C\,{\sc iii} suggests that the degree of ionization, and thus the stellar temperature, of SwSt\,1 has dropped during the last 40 years.  

Shortly redward of this group of three strong carbon lines
appears a small peak in Aller's data which is hardly detectable anymore in the newer spectra. If its tentative identification as C\,{\sc iii} at 5840\,\AA\ is correct, this featurer would contradict the trend of decreasing temperature.

However, in the red region, a significant increase of C\,{\sc ii} 7231 and 7236\,\AA\ lines can be observed. These lines are overestimated with our best fit and contain the emission of both stellar and nebular origin (Fig.\ref{fig:temp}). This is in line with decreasing temperature of the star as well. Nebular emission, which remains unresolved from the stellar component in Fig.\,\ref{fig:aller2}, also contributes to this feature.

\begin{figure}
\includegraphics[angle=270,width=\columnwidth]{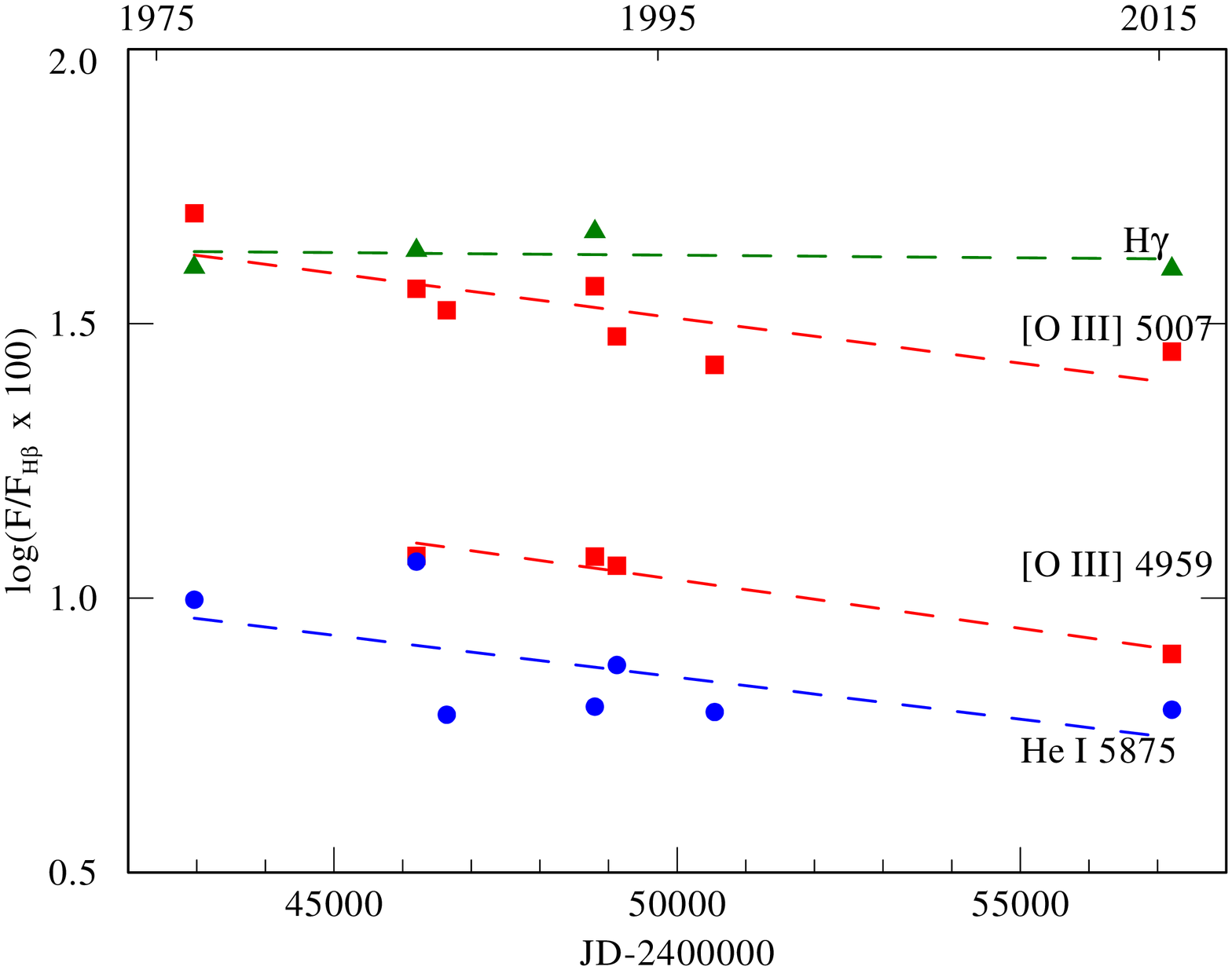}
\caption{Evolution of the flux of selected high excitation lines of the PN SwSt\,1. The dashed lines represent a linear regression to the respective observed values (discrete symbols).}
\label{fig:evol1}
\end{figure}

\begin{figure}
\includegraphics[angle=270,width=\columnwidth]{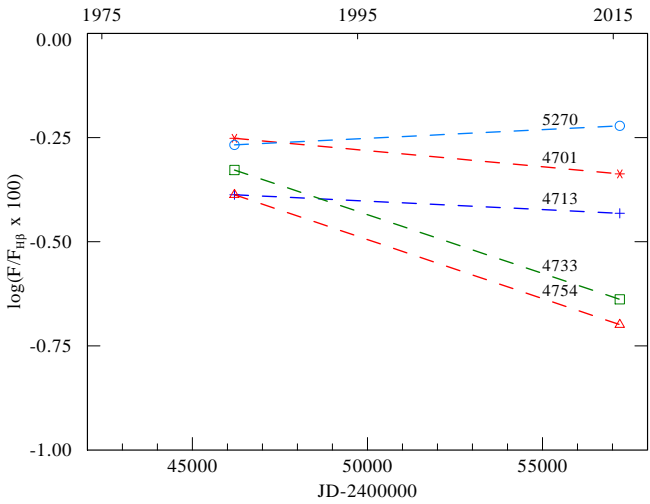}
\caption{Same as Fig.\,\ref{fig:evol1}, but for selected [Fe\,{\sc iii}] lines.}
\label{fig:evol2}
\end{figure}

\begin{figure}
\includegraphics[angle=270,width=\columnwidth]{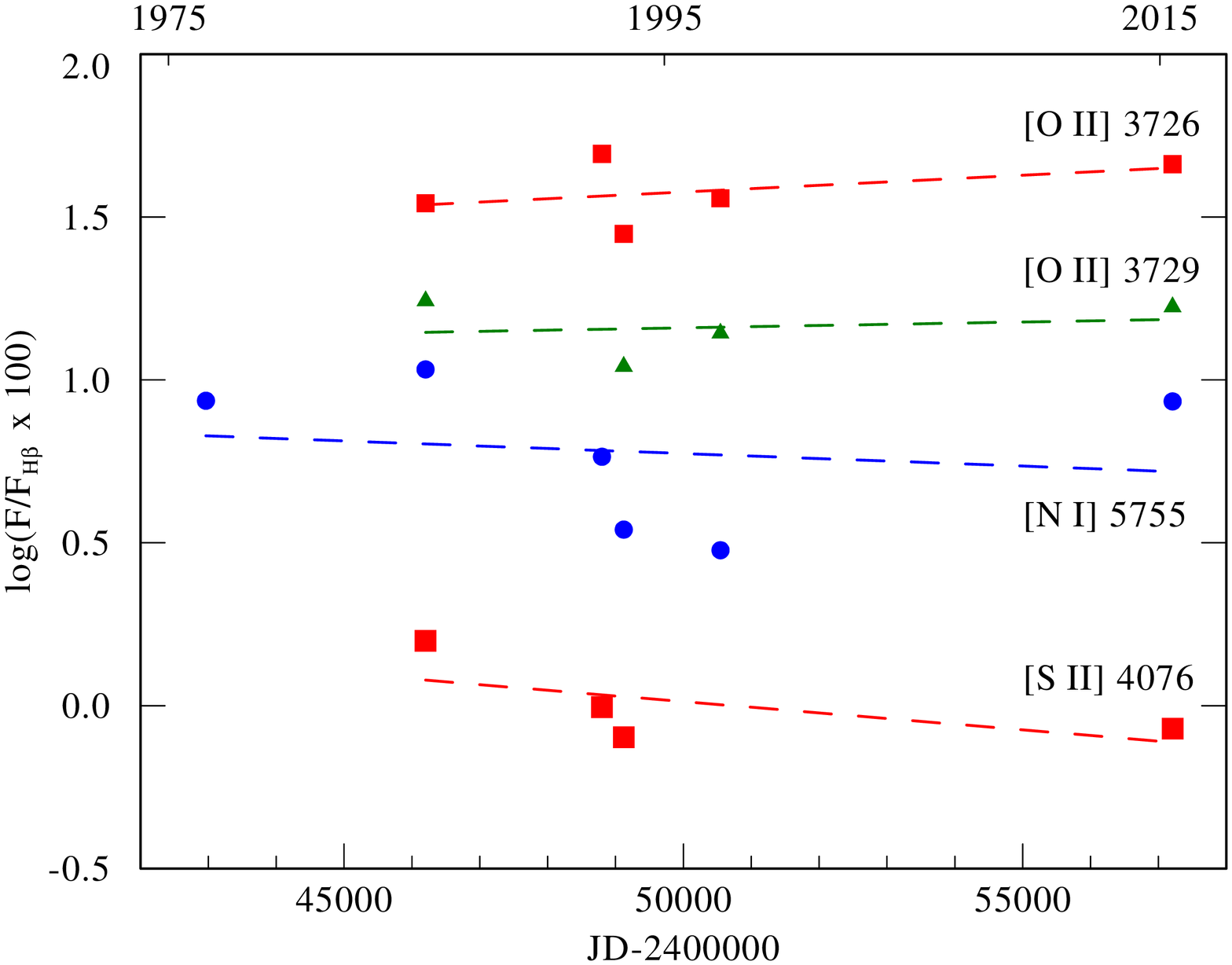}
\caption{Same as Fig.\,\ref{fig:evol1}, but for selected low excitation lines.}
\label{fig:evol3}
\end{figure}

We also fitted the archival spectra of SwSt\,1. Figs.\,\ref{fig:aller1} and \ref{fig:aller3} present the spectrum by \citet{1977JRASC..71...67A} compared with a temperature sequence of models. The best fit to the \citet{1977JRASC..71...67A} spectrum is obtained with the models with $T_*=42 \, \mathrm{kK}$ and $\log R_{\mathrm{t}} = 1.4$. Thus, the comparison with the most recent data indicates a drop of the stellar temperature from about $42.0 \pm 0.5 \, \mathrm{kK}$ to $40.5 \pm 0.5\,\mathrm{kK}$.

\section{Nebular lines and Cloudy modeling}

\begin{table}
\renewcommand{\thetable}{\arabic{table}}
\centering
\caption{Basic parameters adopted and derived from the photoionization models of PN SwSt\,1. The abundances are given by number.} \label{tab:cloudy}
\begin{tabular}{cccc}
\hline
\hline
\\[-1.5ex]
Model     & This paper         & \citetalias{2001MNRAS.328..527D} \\
\\[-1.5ex]
\hline
\\[-1.5ex]
$r_\mathrm{in}$ [pc]  & 0.0015   & 0.0086 \\
$r_\mathrm{out}$ [pc] & 0.0052   & 0.0252 \\
$ \log n_\mathrm{H} \, [\rm cm^{-3}]$ & 5.13 & 4.50 \\
$ \log F (\mathrm{H} \beta) \, [\mathrm{erg} \, \mathrm{cm}^{-2} \, \mathrm{s^{-1}}]$ & -10.06 & -10.1 \\
He/H      & 0.18  & 0.04 \\
log(N/H)  & -3.97 & -4.01 \\
log(O/H)  & -3.33 & -3.85 \\
log(S/H)  & -5.15 & -5.38 \\
log(Cl/H) & -6.74 & \\
log(Ar/H) & -5.39 & \\
log(Fe/H) & -5.4 & \\
$T_\star$ [kK] & $39.2 \pm 0.5$ & $40 \pm 2$ \\
$d$ [kpc] & 2 & 2 \\
\\[-1.5ex]
\hline
\end{tabular}
\end{table}

We used Cloudy v.\,17.01 for photoionization modeling of the nebula \citep{2017RMxAA..53..385F}. Our aim was to derive the temperature evolution of the central star in time using nebular lines. The PoWR model atmospheres provided the incident radiation field. The stellar temperature was a free parameter between 38\,kK and 50\,kK.

We assumed a stellar luminosity of $\log L/\mathrm{L}_{\odot} = 3.27$. We used the $\log \, F (\rm H \alpha$) of $-9.93$ from \citet{2013MNRAS.431....2F}, and dereddened it with $E(B-V)=0.32 \, \mathrm{mag}$ using the extinction law by \citet{1999PASP..111...63F} with the parameter $R_V=3.1$. The nebular line flux ratios were taken from the dereddened VLT spectrum. Abundances were varied to fit the observed line ratios. The basic parameters assumed and derived from the photoionization model are given in Table\,\ref{tab:cloudy}.

\citetalias{2001MNRAS.328..527D} adopted an electron density of $\log (n_\mathrm{e}/\mathrm{cm}^3) = 4.5$ using a diagnostic diagram. \citet{1984MNRAS.206..293F} and \citet{1987MNRAS.227..773D} derived $\log (n_\mathrm{e}/\mathrm{cm}^3) = \rm 5$ using three diagnostic ratios, two of which which were considered as unreliable by \citetalias{2001MNRAS.328..527D}. \citet{2005ApJ...625..368S} derived a density of $\log (n_\mathrm{e}/\mathrm{cm}^3) = 3.94$ from the column density ratio of the excited S\,{\sc iii} fine-structure levels.

We estimate the nebular density using [Fe\,{\sc iii}] line flux ratios. The [Fe\,{\sc iii}] 4702, 4733, 4769, and 4778\,\AA\ to 4658\,\AA\ line ratios indicate $\log (n_\mathrm{e}/\mathrm{cm}^3)$ density between 4.2 and 4.4\,[cm$^{-3}$]. The 5011\,\AA\ and 5270\,\AA\ [Fe\,{\sc iii}]  to 4658\,\AA\ line ratios indicate $\log(n_\mathrm{e})$ of 4.8 and 5.5, respectively. The observed 4881\,\AA\ to 4858\,\AA\ [Fe\,{\sc iii}] line ratio is two times smaller than expected and does not provide a reliable estimate \citep{2017ApJ...841....3L}. We derived the hydrogen density of $\log (n_\mathrm{H}/\mathrm{cm}^3) = 5.13$ for our Cloudy model.

After we derived a satisfactory model, we also attempted to fit the flux ratio of [O\,{\sc iii}] 5007\,\AA\ at different epochs using different $T_{\star}$. The [O\,{\sc iii}] 5007\,\AA\ to $\rm H\beta$ line ratio is a good indicator of the stellar temperature for cool central stars of PNe, assuming that the density and nebular composition did not change between different epochs.

\subsection{Temporal variability of the emission line fluxes}

The PN SwSt\,1 was observed spectroscopically at several epochs (Table\,{\ref{tab:nebula}}). The most pronounced changes can be observed for the [O\,{\sc iii}] 4959 and 5007\,\AA\ lines and the [Fe\,{\sc iii}] lines. Fluxes of high ionization [O\,{\sc iii}] 4959 and 5007\,\AA\ lines decrease with time (Fig.\,\ref{fig:evol1}). The linear fit gives the [O\,{\sc iii}] 5007\,\AA\ to $\mathrm{H}\beta$ flux ratio of $(47 \pm 5) - (0.0014 \pm 0.0005) \, \mathrm{day}^{-1} \times (\mathrm{JD} - 244000)$. The [O\,{\sc iii}] 5007\,\AA\ flux recorded in 1976 was above the linear trend fitted in Fig.\,\ref{fig:evol1}. This may indicate that the drop of the flux was not linear with time. Alternatively, the 1976 [O\,{\sc iii}] 5007\,\AA\ flux was overestimated by about 20\%. The linear fit to the He\,{\sc i} 5875\,\AA\ to H$\beta$ line flux ratio gives  $(10.2 \pm 1.8) - (0.00029 \pm 0.00018) \, \mathrm{day}^{-1} \times (\mathrm{JD} - 244000)$.

\begin{table*}
\renewcommand{\thetable}{\arabic{table}}
\centering
\caption{Modeled and observed nebular lines in the PN SwSt\,1 at different epochs. We used the dereddened fluxes provided by the authors. The numbers denote emission line fluxes normalized to 100 for $\rm{H} \beta$} \label{tab:nebula}
\begin{tabular}{ccccccccccc}
\hline
\hline
\\[-1.5ex]
$\lambda$ [\AA]	&Line&40\,kK&50\,kK &76-07-05	&85-05-16/7	&86-07-13	&92-06-26/8	&93-05-14/5	&	97-04-08&15-07-05	\\
                &   &       &       &   [1]      &   [2]    &   [3]   
&	  [4]   & [5]	& [6]  & [7]	\\
\\[-1.5ex]
\hline
\\[-1.5ex]
3722	&	[S\,{\sc iii}]+H$\kappa$	&	0.4	&	0.7	&		&	3.3	&		&		&		&		&	3.3	\\
3726	&	[O\,{\sc ii}]	&	51.1	&	61.4	&		&	34.9	&		&	49.4	&	28.1	&	36.1	&	45.9	\\
3729	&	[O\,{\sc ii}]	&	22.5	&	27.1	&		&	17.5	&		&		&	11.0	&	13.9	&	16.7	\\
3770	&	H$\iota$	&		&		&		&	2.7	&		&	3.3	&		&		&	3.5	\\
3798	&	H$\theta$	&		&		&		&	3.7	&		&	4.6	&		&		&	4.2	\\
3820	&	He\,{\sc i}	&	0.3	&	0.6	&		&		&		&	0.7	&		&		&	0.5	\\
3835	&	H$\eta$	&		&		&		&	5.2	&		&	6.5	&		&		&	5.7	\\
3868	&	[Ne\,{\sc iii}]	&	0.0	&	0.1	&		&		&		&		&	0.1	&	0.1	&	0.1	\\
3888	&	H$\zeta$	&		&		&		&	10.7	&		&	12.4	&		&		&	10.2	\\
3970	&	H$\epsilon$	&		&		&		&	12.9	&		&	15.3	&		&		&	13.1	\\
4026	&	He\,{\sc i}	&	0.5	&	1.1	&		&		&		&	1.1	&		&		&	0.9	\\
4068	&	[S\,{\sc ii}]	&	5.6	&	1.8	&		&	4.2	&		&	3.8	&	2.0	&	1.1	&	2.6	\\
4076	&	[S\,{\sc ii}]	&	1.8	&	0.6	&		&	1.6	&		&	1.0	&	0.8	&		&	0.8	\\
4101	&	H$\delta$	&		&		&		&	23.2	&		&	25.9	&		&		&	21.5	\\
4267	&	C\,{\sc ii}	&		&		&		&	0.2	&		&	0.7	&		&	0.2	&	0.4	\\
4340	&	H$\gamma$	&		&		&	40.0	&	43.0	&		&	46.5	&		&		&	39.8	\\
4363	&	[O\,{\sc iii}]	&	0.0	&	0.1	&	0.3	&	0.4	&		&	0.4	&	0.3	&	0.2	&	0.2	\\
4368	&	O\,{\sc i}	&	0.0	&	0.0	&		&	0.3	&		&		&		&		&	0.0	\\
4388	&	He\,{\sc i}	&	0.1	&	0.3	&		&	0.5	&		&		&		&		&	0.3	\\
4471	&	He\,{\sc i}	&	1.2	&	2.3	&		&	2.6	&	1.9	&	2.2	&	2.4	&		&	1.9	\\
4657	&	[Fe\,{\sc iii}]	&	8.3	&	8.1	&		&	2.0	&		&		&		&		&	1.2	\\
4686	&	He\,{\sc ii}	&	0.0	&	0.0	&		&		&	0.5	&	0.7	&		&		&	0.7	\\
4701	&	[Fe\,{\sc iii}]	&	3.6	&	3.3	&		&	0.6	&		&		&		&		&	0.5	\\
4713	&	He\,{\sc i}	&	0.2	&	0.3	&		&	0.4	&		&		&		&		&	0.4	\\
4733	&	[Fe\,{\sc iii}]	&	1.6	&	1.3	&		&	0.5	&		&		&		&		&	0.2	\\
4754	&	[Fe\,{\sc iii}]	&	1.5	&	1.5	&		&	0.4	&		&		&		&		&	0.2	\\
4861	&	H$\beta$	&		&		&	100.0	&	100.0	&		&		&		&		&	100.0	\\
4922	&	He\,{\sc i}	&	0.3	&	0.6	&		&	0.8	&		&		&		&		&	0.6	\\
4959	&	[O\,{\sc iii}]	&	1.7	&	12.7	&		&	11.9	&		&	11.9	&	11.4	&		&	7.9	\\
5007	&	[O\,{\sc iii}]	&	5.1	&	38.0	&	50.2	&	36.6	&	33.4	&	37.0	&	30.0	&	26.6	&	28.1	\\
5015	&	He\,{\sc i}	&	0.7	&	1.4	&		&	1.4	&		&		&		&		&	1.3	\\
5192	&	[Ar\,{\sc iii}]	&	0.0	&	0.0	&		&		&		&		&		&		&	0.0	\\
5198	&	[N\,{\sc i}]	&	0.0	&	0.0	&		&		&		&		&		&		&	0.1	\\
5200	&	[N\,{\sc i}]	&	0.0	&	0.0	&		&		&		&		&		&		&	0.0	\\
5270	&	[Fe\,{\sc iii}]	&	7.0	&	7.2	&		&	0.5	&		&		&		&		&	0.6	\\
5517	&	[Cl\,{\sc iii}]	&	0.3	&	0.4	&		&	0.3	&		&	0.1	&		&		&	0.1	\\
5537	&	[Cl\,{\sc iii}]	&	0.3	&	0.5	&		&	0.8	&		&	0.3	&		&		&	0.4	\\
5577	&	[O\,{\sc i}]	&	0.0	&	0.0	&		&		&		&		&	0.2	&		&	0.0	\\
5755	&	[N\,{\sc ii}]	&	0.6	&	0.7	&	8.6	&	10.8	&		&	5.8	&	3.5	&	3.0	&	8.6	\\
5875	&	He\,{\sc i}	&	3.1	&	6.4	&	9.9	&	11.7	&	6.1	&	6.3	&	7.5	&	6.2	&	6.3	\\
6300	&	[O\,{\sc i}]	&	0.5	&	0.1	&		&	1.2	&	0.8	&	1.9	&	1.6	&	1.5	&	1.3	\\
6312	&	[S\,{\sc iii}]	&	0.7	&	1.3	&		&	2.0	&		&	3.1	&	2.8	&	2.5	&	2.7	\\
6363	&	[O\,{\sc i}]	&	0.2	&	0.3	&		&	0.4	&		&	0.5	&	0.5	&		&	0.4	\\
6548	&	[N\,{\sc ii}]	&	34.3	&	27.6	&		&	41.8	&		&	46.4	&	34.4	&		&	42.5	\\
6563	&	H$\alpha$	&		&		&		&	279.7	&		&	285.0	&		&		&	270.9	\\
6584	&	[N\,{\sc ii}]	&	101.1	&	81.4	&		&	127.6	&	102.7	&	149.0	&		&	123.0	&	125.5	\\
6678	&	He\,{\sc i}	&	0.8	&	1.7	&		&	1.9	&		&	1.9	&	19.0	&		&		\\
6717	&	[S\,{\sc ii}]	&	8.4	&		&		&	0.9	&	0.5	&	0.8	&		&	0.7	&		\\
6730	&	[S\,{\sc ii}]	&	14.4	&	3.9	&		&	2.0	&	1.2	&	1.9	&		&	1.6	&		\\
7065	&	He\,{\sc i}	&		&		&		&		&		&	8.1	&		&		&		\\
7135	&	[Ar\,{\sc iii}]	&	3.0	&	7.2	&		&		&		&	12.1	&		&	6.0	&		\\
7325	&	[O\,{\sc ii}]	&		&		&		&		&		&	112.0	&		&		&		\\
\\
\\[-1.5ex]
\hline
\\[-1.5ex]
\end{tabular}

[1] \citet{1984MNRAS.206..293F}
[2] \citet{1987MNRAS.227..773D} 
[3] \citet{1992secg.book.....A}
[4] \citet{2004MNRAS.349.1291E} 
\newline [5] \citetalias{2001MNRAS.328..527D}
[6] \citet{2001A&A...367..983P}
[7] VLT - this paper
\end{table*}

The [O\,{\sc iii}] 5007\,\AA\ line flux was slightly higher in 2015 than in 1997, while the fainter [O\,{\sc iii}] 4959\,\AA\ line continued to decline. This discrepancy can be attributed to the observational uncertainties since both lines originate from the same upper level and have fixed flux ratio.

\citet{1943ApJ....97..194S} present a spectrum of the star in which the [O\,{\sc iii}] 5007\,\AA\ flux appears to be nearly equal to $\rm H\beta$ and stronger than $\rm H\gamma$. This indicates that the [O\,{\sc iii}] 5007\,\AA\ line flux continued to decline at least from around 1940 to 1997. Fluxes of four out of five [Fe\,{\sc iii}] lines also decrease with time (Fig.\,\ref{fig:evol2}), although we have only two data points available for each line. At the same time, low ionization lines do not show a systematic decrease (Fig.\,\ref{fig:evol3}). 

Our analysis of the nebular lines indicates a similar drop of the stellar temperature from about 41.7\,kK around 1976 to about 41\,kK in 1985/86 and 39.2\,kK in 1997 and in 2015. For the derived density, the recombination timescale for $\rm O^{++}$ is less than one year, so this ion should remain in photoionization equilibrium with the radiation field. 

\citet{1987MNRAS.227..773D} attributed the decrease of the [Fe\,{\sc iii}] line fluxes to an increasing ionization of the nebula. However, according to our photoionization models, [Fe\,{\sc iii}] lines weaken for decreasing temperatures of the central star. 

\subsection{Nebular abundances}

[Fe\,{\sc ii}] and [Fe\,{\sc iii}] lines allow for determination of the iron abundance in the gas phase of $\log(\mathrm{Fe}/\mathrm{H}) = -5.4$ by number. This indicates that most of the iron condensed to dust grains, but this factor is somewhat lower than for other PNe \citep{2014ApJ...784..173D}.

The observed spectrum shows C\,{\sc ii}, N\,{\sc ii}, O\,{\sc i}, and O\,{\sc ii} lines stronger than the model by a factor of $8.5-18$. We attribute this to the so-called abundance discrepancy, often observed in PNe \citep[and references therein]{2018MNRAS.480.4589}.

\section{Discussion}

\begin{figure}
\includegraphics[width=\columnwidth]{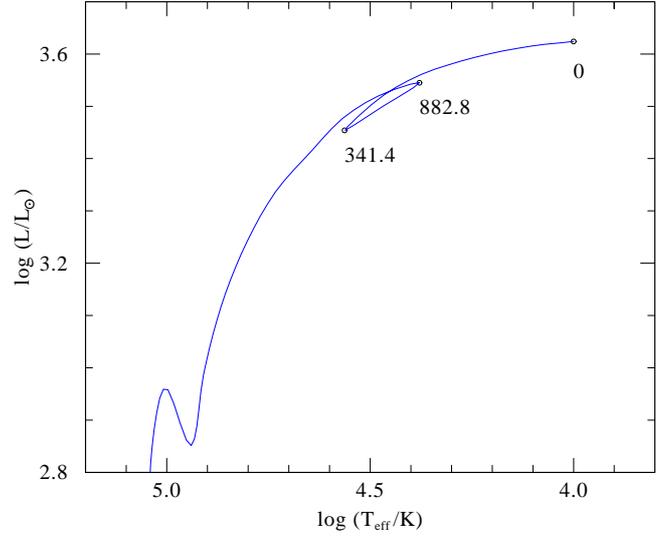}
\caption{The evolution of a He-burning post-AGB star. The dashed line marks the track for the LTP for a final mass of $\rm 0.558 \, \mathrm{M}_{\odot}$ \citep{1994ApJS...92..125V}. The labels indicate the time in yr at the start of the late thermal pulse and at the second minimum of the stellar temperature.}
\label{fig:vassiliadis}
\end{figure}

Single star evolutionary models predict a gradual heating of the star at constant luminosity after having left the AGB. After the star has reached its maximum temperature, the stellar luminosity drops while the star enters the WD cooling sequence. However, if either the H or the He shell becomes activated again, the released energy forces the star to expand and cool down. The LTP object FG\,Sge evolved from a hot H rich post-AGB star to a cool giant for over 120 years \citep{2006A&A...459..885J}. For the VLTP objects V605\,Aql and V4334\,Sgr a transition from WD to a cool giant took only a few years.

\citet{2007MNRAS.380..763M} found that a massive post-AGB star (final mass $M_\mathrm{f} \gtrsim 0.6 \, \mathrm{M}_{\odot}$, corresponding to an initial mass of $M_i \gtrsim 3\,\mathrm{M}_{\odot}$) does not expand significantly as the result of the violent H burning which takes place during the VLTP. The energy released as a result of the hydrogen ingestion is lower than the gravitational binding energy of the envelope. Thus, the VLTP models of massive post-AGB stars predict only one return to the AGB domain, powered by the helium flash.

A massive post-AGB star has only a very tiny envelope. \citet{1970AcA....20...47P} shows that a post-AGB star with a small H envelope mass does not reach the giant branch as a result of a thermal pulse. Instead, it performs a smaller loop in the HR diagram. Moreover, \citet{1986ApJ...307..659W} argue that in the case of an AGB star capable to produce a nucleus more massive than $0.86\,\mathrm{M}_{\odot}$, energy generated at the bottom of the convective envelope during a helium flash exceeds the Eddington luminosity. The instability can strip off the envelope, and the object evolves to a H-deficient central star of a PN. \citet{2012A&A...542A...1L} confirmed that super-Eddington luminosity develops at the bottom of the convective envelope for massive post-AGB stars in the late thermally-pulsing phase.

The evolution of stellar and nebular spectra of SwSt\,1 indicates a possible drop of the stellar temperature. The observations suggest that the decrease of the temperature stopped at some time between 1997 and 2015. The star may now be heating again. This suggests that the star has completed a loop in the HR diagram, due to reactivation of He shell burning, but without reaching the red giant domain.

The star was already H-poor in 1895 \citepalias{2001MNRAS.328..527D}, and hence it must have already been burning helium at that time. No photometric variability was reported, suggesting that the optical brightness of the object did not change significantly. 

The evolution of the star appears to be very fast. The dynamical age of the nebula is only 290 years for the assumed distance of 2\,kpc.  The derived age assumes that the expansion velocity corresponds to the outer edge of the ionized nebula. Also, there could be a neutral part of the ejecta around the ionized nebula which could be older. However, the ionized nebula was ejected by a cool giant that has left the AGB $\lesssim 290 \, \rm yr$ ago.

One possible evolutionary scenario for the central star of PN SwSt\,1 is a VLTP experienced by a massive post-AGB star. A $0.833 \mathrm{M}_{\odot}$ model in \citet{2016A&A...588A..25M} reaches the WD cooling sequence in less than 100\,yr. The born-again time, from the start of the VLTP until the return to the giant region of the HR diagram, is only 65 years for a $0.741 \mathrm{M}_{\odot}$ core \citep{2006A&A...449..313M}. 

A high mass of the star is supported by its high luminosity. A star of final mass $M_\mathrm{f} \gtrsim 0.6 \, \mathrm{M}_{\odot}$ should exhibit a luminosity of about $ \log L/\mathrm{L}_{\odot} \approx 3.9$ \citep{2007MNRAS.380..763M}, which would be consistent with the distance of 3\,kpc.

The N abundance, however, indicates that SwSt\,1 is a type\mbox{}II PN. Such PNe originate from progenitors with $M_\mathrm{i} \lesssim 2.4 \, \mathrm{M}_{\odot}$ \citep{1994MNRAS.271..257K}.

The VLTP consumes hydrogen in the stellar atmosphere. This results in $\beta_\mathrm{N}$ of more than 20\% and $\beta_\mathrm{O}$ more than 1\,\% \citep{2006A&A...449..313M}, much higher than we derived for the central star of the PN SwSt\,1 (Table \ref{tab:powr}).

An alternative scenario would be an LTP experienced by a massive post-AGB star. Fig.\,\ref{fig:vassiliadis} presents an evolutionary model for a LTP by \citet{1994ApJS...92..125V}. The average heating rate in the part of the track where the star decreases in temperature is $\dot{T_*} = -23 \, \mathrm{K} \, \mathrm{yr^{-1}}$. We derived a heating rate of $\dot{T_*} = -88 \, \mathrm{K} \, \mathrm{yr^{-1}}$ for our star between 1977 and 1993, which would indicate a more massive post-AGB star. An LTP in a massive post-AGB star can result in a hydrogen-poor atmosphere.

It is difficult to conclude if interaction with a companion or planet(s) influenced the evolution of the central star of SwSt\,1. The star does not reveal photometric variability due to a close companion \citep{2010MNRAS.406..626H}. High resolution observations (the SALT, Mercator, and VLT telescopes) at three different dates do not show radial velocity differences higher than about $10\,\mathrm{km \, s^{-1}}$. A binary with an orbit plane perpendicular to the line of sight, a merger scenario, or interaction with a planet at late AGB phase cannot be ruled out, however.

SwSt\,1 appears to present a special evolutionary channel for the formation of [WC] stars. \citetalias{1998AA...330..265L} noted that SwSt\,1 does not follow the evolutionary track for other [WC] stars in the $R_\mathrm{t} - T_{\star}$ diagram since it has a lower mass loss rate. Further monitoring of the object is needed to constrain its evolutionary status, and thus to better understand the processes involved in late thermal pulses and the evolution of massive post-AGB stars.

\section{Acknowledgements}

The Polish-German collaboration was supported by the Deutscher Akademischer Austauschdienst (DAAD) and Ministry of Science and Higher Education (MSHE) of the Republic of Poland. MH and KB acknowledge financial support from National Science Centre, Poland, grant No. 2016/23/B/ST9/01653, the MSHE for granting funds for the Polish contribution to the International LOFAR Telescope (MSHE decision no. DIR/WK/2016/2017/05-1) and for maintenance of the LOFAR PL-612 Baldy (MSHE decision no.~59/E-383/SPUB/SP/2019.1). Polish participation in SALT is funded by grant No. MSHE DIR/WK/2016/07. The paper is partially based on observations collected at the European Southern Observatory under ESO programmes 093.D-0182(A) and 095.D-0953(A). Some of the observations reported in this paper were obtained with the Southern African Large Telescope (SALT) under programme 2015-1-SCI-031 (PI M. Hajduk). 

\section{Data Availability Statement}

The data used in this paper are available in the ESO Science Archive Facility at http://archive.eso.org/cms.html and in the SALT Data Archive at https://ssda.saao.ac.za/.




\clearpage

\appendix
\section{Nebular lines in the VLT spectrum of SwSt\,1} \label{tab:vlt}
\begin{supertabular}{ccccc}
\hline
\hline
\\[-1.5ex]
 $\lambda$ & FWHM	& F/F(H$\beta$)	& I/I(H$\beta$) & ion \\
 \r{A}   & [km/s] &               &               &    \\
\\[-1.5ex]
\hline
\\[-1.5ex]
3296.77	&	29.44	&	0.036	&	0.059	&	He\,{\sc i}	\\
3342.85	&	29.62	&	0.091	&	0.143	&	[Cl\,{\sc iii}]	\\
3353.21	&	28.50	&	0.036	&	0.057	&	[Cl\,{\sc iii}]	\\
3354.55	&	32.68	&	0.073	&	0.115	&	He\,{\sc i}	\\
3371.30	&	42.09	&	0.022	&	0.034	&	N\,{\sc ii}? [Fe\,{\sc iii}]?	\\
3423.07	&	97.64	&	0.028	&	0.043	&	N\,{\sc ii}? [Fe\,{\sc iii}]?	\\
3427.57	&	54.96	&	0.081	&	0.125	&	N\,{\sc ii}?	\\
3444.23	&	87.72	&	0.139	&	0.211	&	He\,{\sc i} blend?	\\
3447.59	&	39.53	&	0.160	&	0.243	&	He\,{\sc i}	\\
3461.01	&	20.36	&	0.009	&	0.013	&	He\,{\sc i}	\\
3498.64	&	42.86	&	0.041	&	0.061	&	He\,{\sc i}	\\
3512.51	&	38.57	&	0.042	&	0.063	&	He\,{\sc i}	\\
3530.49	&	39.62	&	0.053	&	0.078	&	He\,{\sc i}	\\
3554.41	&	38.24	&	0.070	&	0.103	&	He\,{\sc i}	\\
3587.27	&	39.03	&	0.099	&	0.144	&	He\,{\sc i}	\\
3613.64	&	35.37	&	0.191	&	0.275	&	He\,{\sc i}	\\
3634.23	&	38.19	&	0.146	&	0.208	&	He\,{\sc i}	\\
3656.11	&	36.07	&	0.031	&	0.044	&	H\,{\sc i}	\\
3656.66	&	29.50	&	0.036	&	0.051	&	H\,{\sc i}	\\
3657.27	&	30.11	&	0.047	&	0.066	&	H\,{\sc i}	\\
3657.92	&	30.54	&	0.063	&	0.090	&	H\,{\sc i}	\\
3658.64	&	31.56	&	0.080	&	0.114	&	H\,{\sc i}	\\
3659.42	&	33.55	&	0.101	&	0.143	&	H\,{\sc i}	\\
3660.28	&	34.86	&	0.125	&	0.177	&	H\,{\sc i}	\\
3661.22	&	38.13	&	0.165	&	0.234	&	H\,{\sc i}	\\
3662.26	&	37.99	&	0.182	&	0.257	&	H\,{\sc i}	\\
3663.40	&	38.38	&	0.203	&	0.287	&	H\,{\sc i}	\\
3664.68	&	40.18	&	0.244	&	0.345	&	H\,{\sc i}	\\
3666.09	&	38.96	&	0.251	&	0.355	&	H\,{\sc i}	\\
3667.68	&	38.77	&	0.280	&	0.396	&	H\,{\sc i}	\\
3669.46	&	38.76	&	0.318	&	0.449	&	H\,{\sc i}	\\
3671.48	&	40.03	&	0.367	&	0.518	&	H\,{\sc i}	\\
3673.76	&	39.98	&	0.410	&	0.578	&	H\,{\sc i}	\\
3676.36	&	40.79	&	0.471	&	0.665	&	H\,{\sc i}	\\
3679.35	&	40.63	&	0.534	&	0.752	&	H\,{\sc i}	\\
3682.81	&	40.61	&	0.599	&	0.842	&	H\,{\sc i}	\\
3686.83	&	40.02	&	0.631	&	0.887	&	H\,{\sc i}	\\
3691.55	&	39.41	&	0.714	&	1.001	&	H\,{\sc i}	\\
3697.15	&	39.80	&	0.822	&	1.151	&	H\,{\sc i}	\\
3703.85	&	42.63	&	1.077	&	1.506	&	H\,{\sc i}	\\
3705.00	&	39.75	&	0.234	&	0.327	&	He\,{\sc i}	\\
3711.97	&	40.00	&	1.074	&	1.498	&	H\,{\sc i}	\\
3721.94	&	44.95	&	2.405	&	3.344	&	H\,{\sc i}+[Si\,{\sc ii}]?	\\
3726.03	&	38.05	&	33.074	&	45.927	&	[O\,{\sc ii}]	\\
3728.81	&	39.39	&	12.066	&	16.741	&	[O\,{\sc ii}]	\\
3734.37	&	39.99	&	1.511	&	2.094	&	H\,{\sc i}	\\
3750.15	&	40.20	&	2.008	&	2.768	&	H\,{\sc i}	\\
3768.74	&	33.41	&	0.009	&	0.013	&	He\,{\sc i}	\\
3770.63	&	40.27	&	2.535	&	3.474	&	H\,{\sc i}	\\
3784.82	&	34.20	&	0.010	&	0.014	&	He\,{\sc i}	\\
3797.90	&	39.89	&	3.054	&	4.152	&	H\,{\sc i}	\\
3805.74	&	97.05	&	0.041	&	0.055	&	He\,{\sc i}	\\
3819.60	&	36.91	&	0.363	&	0.490	&	He\,{\sc i}	\\
3829.80	&	45.83	&	0.012	&	0.016	&	N\,{\sc ii}? MgI?	\\
3831.73	&	42.13	&	0.017	&	0.024	&	C\,{\sc ii}?	\\
3833.57	&	39.29	&	0.021	&	0.028	&	He\,{\sc i}	\\
3835.38	&	39.45	&	4.219	&	5.674	&	H\,{\sc i}	\\
3838.37	&	34.95	&	0.025	&	0.033	&	N\,{\sc ii}?	\\
3856.02	&	35.45	&	0.105	&	0.140	&	Si\,{\sc ii}	\\
3862.60	&	31.23	&	0.037	&	0.049	&	Si\,{\sc ii}	\\
3867.48	&	30.54	&	0.026	&	0.035	&	He\,{\sc i}	\\
3868.76	&	39.85	&	0.098	&	0.130	&	[Ne\,{\sc iii}]	\\
3871.69	&	35.88	&	0.030	&	0.040	&	He\,{\sc i}?	\\
3889.05	&	42.32	&	7.684	&	10.174	&	H\,{\sc i}	\\
3918.98	&	37.16	&	0.096	&	0.126	&	C\,{\sc ii}?	\\
3920.69	&	37.18	&	0.197	&	0.258	&	C\,{\sc ii}?	\\
3926.41	&	37.73	&	0.044	&	0.058	&	He\,{\sc i}	\\
3964.73	&	35.11	&	0.412	&	0.533	&	He\,{\sc i}	\\
3970.07	&	37.21	&	10.101	&	13.071	&	H\,{\sc i}	\\
3995.00	&	26.62	&	0.016	&	0.021	&	N\,{\sc ii}	\\
4008.35	&	36.81	&	0.038	&	0.048	&	[Fe\,{\sc iii}]?	\\
4009.26	&	32.68	&	0.056	&	0.072	&	He\,{\sc i}	\\
4026.20	&	36.56	&	0.719	&	0.915	&	He\,{\sc i}	\\
4068.60	&	33.43	&	2.094	&	2.636	&	[S\,{\sc ii}]	\\
4076.35	&	35.25	&	0.675	&	0.847	&	[S\,{\sc ii}]	\\
4079.70	&	32.62	&	0.009	&	0.011	&	[Fe\,{\sc iii}]?	\\
4101.73	&	36.90	&	17.260	&	21.533	&	H\,{\sc i}	\\
4120.99	&	32.28	&	0.058	&	0.073	&	He\,{\sc i}	\\
4131.73	&	32.89	&	0.019	&	0.023	&	N\,{\sc ii}?	\\
4143.76	&	36.26	&	0.112	&	0.138	&	He\,{\sc i}+[Fe\,{\sc ii}]	\\
4168.97	&	41.88	&	0.025	&	0.031	&	He\,{\sc i}	\\
4243.97	&	40.22	&	0.021	&	0.025	&	[Fe\,{\sc ii}]	\\
4267.00	&	48.90	&	0.197	&	0.235	&	C\,{\sc ii}?	\\
4276.83	&	49.79	&	0.025	&	0.029	&	[Fe\,{\sc ii}]	\\
4287.39	&	39.85	&	0.056	&	0.066	&	[Fe\,{\sc ii}]	\\
4349.43	&	37.91	&	0.027	&	0.031	&	O\,{\sc ii}	\\
4359.33	&	39.53	&	0.046	&	0.054	&	[Fe\,{\sc ii}]	\\
4363.21	&	30.14	&	0.186	&	0.216	&	[O\,{\sc iii}]	\\
4387.93	&	41.85	&	0.285	&	0.329	&	He\,{\sc i}	\\
4408.54	&	62.07	&	0.026	&	0.030	&	Si\,{\sc iii}?	\\
4413.78	&	51.47	&	0.048	&	0.055	&	[Fe\,{\sc ii}]	\\
4414.90	&	49.92	&	0.077	&	0.088	&	O\,{\sc ii}?	\\
4416.27	&	72.90	&	0.073	&	0.083	&	[Fe\,{\sc ii}]	\\
4416.97	&	32.56	&	0.021	&	0.024	&	O\,{\sc ii}?	\\
4437.55	&	35.69	&	0.038	&	0.043	&	He\,{\sc i}	\\
4447.03	&	29.84	&	0.026	&	0.030	&	N\,{\sc ii}	\\
4452.10	&	41.71	&	0.023	&	0.026	&	[Fe\,{\sc ii}]	\\
4471.50	&	35.66	&	1.706	&	1.921	&	He\,{\sc i}	\\
4638.92	&	36.86	&	0.011	&	0.012	&	C\,{\sc ii}??	\\
4658.05	&	38.48	&	1.086	&	1.157	&	[Fe\,{\sc iii}]	\\
4664.08	&	128.34	&	0.248	&	0.264	&	N\,{\sc ii}	\\
4666.94	&	33.93	&	0.049	&	0.052	&	[Fe\,{\sc iii}]	\\
4685.68	&	160.70	&	0.638	&	0.674	&	He\,{\sc ii}	\\
4699.22	&	33.67	&	0.022	&	0.023	&	O\,{\sc ii}?	\\
4701.53	&	35.92	&	0.434	&	0.456	&	[Fe\,{\sc iii}]	\\
4705.35	&	36.33	&	0.027	&	0.029	&	O\,{\sc ii}	\\
4710.01	&	28.82	&	0.007	&	0.007	&	O\,{\sc ii}	\\
4713.20	&	34.96	&	0.350	&	0.367	&	He\,{\sc i}	\\
4728.07	&	33.53	&	0.009	&	0.010	&	[Fe\,{\sc ii}]	\\
4733.91	&	35.25	&	0.220	&	0.229	&	[Fe\,{\sc iii}]	\\
4754.69	&	37.58	&	0.196	&	0.203	&	[Fe\,{\sc iii}]	\\
4769.43	&	35.59	&	0.152	&	0.156	&	[Fe\,{\sc iii}]	\\
4773.47	&	188.88	&	0.013	&	0.013	&	[Fe\,{\sc ii}]	\\
4777.68	&	34.84	&	0.105	&	0.108	&	[Fe\,{\sc iii}]	\\
4779.72	&	44.38	&	0.031	&	0.032	&	N\,{\sc ii}?	\\
4788.13	&	29.72	&	0.018	&	0.019	&	N\,{\sc ii}	\\
4789.45	&	46.41	&	0.030	&	0.031	&	F\,{\sc ii}?	\\
4793.65	&	32.69	&	0.007	&	0.007	&	N\,{\sc ii}?	\\
4798.27	&	53.36	&	0.008	&	0.008	&	[Fe\,{\sc ii}]	\\
4803.29	&	32.83	&	0.015	&	0.015	&	N\,{\sc ii}?	\\
4810.31	&	132.24	&	0.013	&	0.013	&	N\,{\sc ii}?	\\
4810.31	&	26.29	&	0.003	&	0.004	&	N\,{\sc ii}?	\\
4814.53	&	46.25	&	0.037	&	0.037	&	[Fe\,{\sc ii}]	\\
4815.62	&	38.57	&	0.024	&	0.024	&	N\,{\sc ii}?	\\
4852.73	&	57.66	&	0.015	&	0.015	&	[Fe\,{\sc ii}]	\\
4874.48	&	45.71	&	0.007	&	0.007	&	[Fe\,{\sc ii}]	\\
4874.48	&	8.78	&	0.001	&	0.001	&	[Fe\,{\sc ii}]	\\
4881.07	&	40.89	&	0.406	&	0.403	&	[Fe\,{\sc iii}]	\\
4889.62	&	44.19	&	0.027	&	0.027	&	[Fe\,{\sc ii}]	\\
4895.11	&	38.60	&	0.034	&	0.034	&	N\,{\sc ii}?	\\
4905.34	&	55.82	&	0.026	&	0.026	&	[Fe\,{\sc ii}]	\\
4921.93	&	34.53	&	0.608	&	0.597	&	He\,{\sc i}	\\
4930.44	&	44.40	&	0.099	&	0.097	&	[Fe\,{\sc iii}]?	\\
4921.93	&	34.52	&	0.607	&	0.596	&	He\,{\sc i}	\\
4947.37	&	59.08	&	0.006	&	0.006	&	[Fe\,{\sc ii}]	\\
4950.74	&	43.82	&	0.006	&	0.006	&	[Fe\,{\sc ii}]	\\
4958.91	&	30.39	&	8.153	&	7.909	&	[O\,{\sc iii}]	\\
4973.39	&	66.36	&	0.010	&	0.010	&	[Fe\,{\sc ii}]	\\
4980.09	&	38.56	&	0.018	&	0.017	&	O\,{\sc i}?	\\
4987.29	&	39.35	&	0.098	&	0.094	&	[Fe\,{\sc iii}]	\\
4994.37	&	33.27	&	0.061	&	0.059	&	N\,{\sc ii}	\\
5001.13	&	40.31	&	0.033	&	0.032	&	N\,{\sc ii}?	\\
5002.70	&	30.29	&	0.008	&	0.008	&	N\,{\sc ii}?	\\
5006.84	&	31.93	&	29.424	&	28.117	&	[O\,{\sc iii}]	\\
5010.62	&	46.27	&	0.051	&	0.049	&	N\,{\sc ii}?	\\
5011.41	&	34.64	&	0.236	&	0.226	&	[Fe\,{\sc iii}]	\\
5015.70	&	33.42	&	1.327	&	1.265	&	He\,{\sc i}	\\
5032.13	&	48.12	&	0.038	&	0.036	&	C\,{\sc ii}	\\
5035.94	&	40.14	&	0.035	&	0.033	&	C\,{\sc ii}	\\
5041.03	&	34.24	&	0.084	&	0.080	&	Si\,{\sc ii}	\\
5045.10	&	34.54	&	0.039	&	0.036	&	N\,{\sc ii}?	\\
5047.74	&	35.93	&	0.139	&	0.131	&	He\,{\sc i}	\\
5055.98	&	37.34	&	0.238	&	0.224	&	Si\,{\sc ii}	\\
5084.85	&	38.48	&	0.048	&	0.045	&	[Fe\,{\sc iii}]	\\
5111.63	&	53.06	&	0.018	&	0.017	&	[Fe\,{\sc ii}]	\\
5121.83	&	74.05	&	0.028	&	0.026	&	C\,{\sc ii}	\\
5158.79	&	71.48	&	0.125	&	0.114	&	[Fe\,{\sc ii}]	\\
5191.82	&	46.66	&	0.011	&	0.010	&	[Ar\,{\sc iii}]	\\
5197.90	&	60.03	&	0.088	&	0.080	&	[N\,{\sc i}]	\\
5200.26	&	68.66	&	0.055	&	0.050	&	[N\,{\sc i}]	\\
5220.08	&	40.06	&	0.008	&	0.007	&	[Fe\,{\sc ii}]	\\
5261.62	&	43.09	&	0.045	&	0.039	&	[Fe\,{\sc ii}]	\\
5270.57	&	0.00	&	0.677	&	0.598	&	[Fe\,{\sc iii}]	\\
5299.05	&	40.09	&	0.049	&	0.043	&	O\,{\sc i}	\\
5333.65	&	74.20	&	0.026	&	0.023	&	[Fe\,{\sc ii}]	\\
5376.45	&	48.01	&	0.024	&	0.021	&	[Fe\,{\sc ii}]	\\
5411.98	&	41.31	&	0.103	&	0.088	&	[Fe\,{\sc iii}]	\\
5433.13	&	44.17	&	0.016	&	0.013	&	[Fe\,{\sc ii}]	\\
5452.07	&	32.34	&	0.008	&	0.006	&	N\,{\sc ii}	\\
5454.22	&	49.43	&	0.027	&	0.023	&	N\,{\sc ii}	\\
5480.06	&	39.08	&	0.015	&	0.013	&	N\,{\sc ii}	\\
5495.67	&	34.13	&	0.021	&	0.018	&	N\,{\sc ii}	\\
5512.78	&	40.73	&	0.044	&	0.037	&	O\,{\sc ii}?	\\
5517.72	&	37.40	&	0.134	&	0.111	&	[Cl\,{\sc iii}]	\\
5527.34	&	42.11	&	0.014	&	0.012	&	[Fe\,{\sc ii}]	\\
5537.89	&	34.56	&	0.500	&	0.413	&	[Cl\,{\sc iii}]	\\
5551.92	&	41.31	&	0.011	&	0.009	&	[Fe\,{\sc ii}]?	\\
5555.01	&	45.80	&	0.041	&	0.034	&	O\,{\sc ii}?	\\
5577.34	&	41.06	&	0.040	&	0.033	&	[O\,{\sc i}] sky?	\\
5676.02	&	31.43	&	0.036	&	0.028	&	N\,{\sc ii}	\\
5679.56	&	33.52	&	0.063	&	0.051	&	N\,{\sc ii}	\\
5710.77	&	35.45	&	0.021	&	0.016	&	N\,{\sc ii}	\\
5739.73	&	37.29	&	0.012	&	0.010	&	Si\,{\sc iii}	\\
5747.19	&	39.47	&	0.011	&	0.008	&	N\,{\sc ii}?	\\
5754.60	&	29.37	&	10.926	&	8.588	&	[N\,{\sc ii}]	\\
5857.27	&	89.88	&	0.032	&	0.025	&	He\,{\sc ii}	\\
5875.64	&	31.67	&	8.171	&	6.263	&	He\,{\sc i}	\\
5927.81	&	33.31	&	0.037	&	0.028	&	N\,{\sc ii}	\\
5931.78	&	33.70	&	0.048	&	0.036	&	N\,{\sc ii}	\\
5940.24	&	34.00	&	0.026	&	0.019	&	N\,{\sc ii}	\\
5941.65	&	35.88	&	0.038	&	0.029	&	N\,{\sc ii}	\\
5952.39	&	38.45	&	0.017	&	0.013	&	N\,{\sc ii}	\\
5957.56	&	37.60	&	0.043	&	0.032	&	Si\,{\sc ii}	\\
5958.58	&	37.65	&	0.039	&	0.030	&	O\,{\sc i}	\\
5978.98	&	35.95	&	0.084	&	0.063	&	Si\,{\sc ii}	\\
6019.89	&	31.42	&	0.003	&	0.002	&	C\,{\sc i}?	\\
6021.38	&	27.99	&	0.002	&	0.001	&	C\,{\sc i}?	\\
6036.70	&	116.07	&	0.028	&	0.021	&	He\,{\sc ii}	\\
6046.44	&	38.96	&	0.142	&	0.105	&	O\,{\sc i}	\\
6074.10	&	89.35	&	0.014	&	0.010	&	He\,{\sc ii}	\\
6096.16	&	40.85	&	0.008	&	0.006	&	Ne\,{\sc i} ?	\\
6143.06	&	31.51	&	0.006	&	0.004	&	Ne\,{\sc i} ?	\\
6234.36	&	133.98	&	0.019	&	0.014	&	He\,{\sc ii}?	\\
6257.18	&	46.81	&	0.009	&	0.007	&	C\,{\sc ii}	\\
6259.56	&	38.87	&	0.013	&	0.009	&	C\,{\sc ii}	\\
6266.50	&	21.62	&	0.001	&	0.001	&	Ne\,{\sc i} ?	\\
6300.34	&	33.57	&	1.820	&	1.285	&	[O\,{\sc i}]	\\
6312.06	&	28.10	&	3.787	&	2.668	&	[Si\,{\sc ii}]	\\
6334.43	&	46.60	&	0.007	&	0.005	&	N\,{\sc ii}?	\\
6347.09	&	34.84	&	0.223	&	0.156	&	Si\,{\sc ii}	\\
6363.78	&	39.05	&	0.635	&	0.444	&	[O\,{\sc i}]	\\
6371.38	&	32.82	&	0.088	&	0.062	&	Si\,{\sc ii}	\\
6402.25	&	31.39	&	0.018	&	0.012	&	Ne\,{\sc i}?	\\
6460.83	&	125.20	&	0.072	&	0.049	&	Ne\,{\sc ii}?	\\
6482.05	&	29.25	&	0.037	&	0.025	&	N\,{\sc ii}	\\
6516.15	&	107.10	&	0.124	&	0.084	&	N\,{\sc ii}	\\
6527.23	&	43.08	&	0.046	&	0.031	&	N\,{\sc ii}	\\
6548.05	&	37.89	&	62.958	&	42.537	&	[N\,{\sc ii}]	\\
6563.22	&	40.49	&	401.935	&	270.875	&	H\,{\sc i}	\\
6578.01	&	36.98	&	0.768	&	0.516	&	C\,{\sc ii}	\\
6583.45	&	38.13	&	186.902	&	125.518	&	[N\,{\sc ii}]	\\
6610.56	&	34.03	&	0.017	&	0.011	&	N\,{\sc ii}	\\
\\[-1.5ex]
\hline
\end{supertabular}

\bsp	
\label{lastpage}
\end{document}